\documentclass[]{aastex631}

\shorttitle{A need to revise stellar opacities from asteroseismology}
\shortauthors{Daszy\'nska-Daszkiewicz et al.}
\graphicspath{{./}{figures/}}

\begin{document}

%\title{A need to revise stellar opacities from asteroseismology of double-mode radial $\delta$ Scuti stars}
\title{A need to revise stellar opacities from asteroseismology of $\delta$ Scuti stars}

\correspondingauthor{Jadwiga Daszy\'nska-Daszkiewicz}
\email{daszynska@astro.uni.wroc.pl}

%\author[0000-0001-9704-6408]{Jadwiga Daszy\'nska-Daszkiewicz}
\author[0000-0001-9704-6408]{Jadwiga Daszy\'nska-Daszkiewicz}
\affiliation{University of Wroc{\l}aw, Faculty of Physics and Astronomy \\
Astronomical Institute, ul. Kopernika 11\\
PL-51-622 Wroc{\l}aw, Poland}

\author[0000-0003-3476-8483]{Przemys{\l}aw Walczak}
\affiliation{University of Wroc{\l}aw, Faculty of Physics and Astronomy \\
	Astronomical Institute, ul. Kopernika 11\\
	PL-51-622 Wroc{\l}aw, Poland}

\author[0000-0002-5481-3512]{Alexey Pamyatnykh}
\affiliation{Nicolaus Copernicus Astronomical Center, Polish Academy of Sciences\\
    ul. Bartycka 18, PL-00-716 Warsaw, Poland}

\author[0000-0002-2393-8427]{Wojciech Szewczuk}
\affiliation{University of Wroc{\l}aw, Faculty of Physics and Astronomy \\
	Astronomical Institute, ul. Kopernika 11\\
	PL-51-622 Wroc{\l}aw, Poland}

\author{Wojciech Niewiadomski}
\affiliation{University of Wroc{\l}aw, Faculty of Physics and Astronomy \\
	Astronomical Institute, ul. Kopernika 11\\
	PL-51-622 Wroc{\l}aw, Poland}

\begin{abstract}
We construct seismic models of the four double-mode radial  $\delta$ Scuti stars adopting
opacities from three databases:  OPAL, OP and OPLIB.
The aim is to find the models that fit the observed frequencies of the two radial modes
and have the effective temperature and luminosity consistent with the observed values.
Using the Bayesian analysis based on Monte Carlo simulations, we obtain that
only the OPAL seismic models  are caught within the observed error box in the HR diagram.
Seismic models computed with the OP and OPLIB data are much cooler and less luminous.
By including the relative amplitude of the bolometric flux variations (the so-called parameter $f$)
into these simulations, we constrain the efficiency of convection in the envelopes,
described by the mixing length parameter $\alpha_{\rm MLT}$. We get $\alpha_{\rm MLT}\approx 0.5$ for
BP Peg, AE UMa and RV Ari (Population I stars) and $\alpha_{\rm MLT}\approx 1.0$ for SX Phe
(Population II star). For all the stars, overshooting from the convective core seems inefficient.
A similar effect of opacity should occur also for classical Cepheids or RR Lyr stars that are used as standard candles
to measure the universe. %  Thus, the value of $\alpha_{\rm MLT}$ has to be constrained for each star individually.
\end{abstract}

%%\keywords{Classical Novae (251) --- Ultraviolet astronomy(1736) --- History of astronomy(1868) --- Interdisciplinary astronomy(804)}

\section{Introduction} \label{sec:intro}
Asteroseismology provides the most stringent constraints  on the theory of
stellar structure and evolution. It also offers a test of microphysics data, in particular
stellar opacities, which are among the major and still uncertain components of modern astrophysics.
%This fact was already highlighted by Eddington in 1926.

The opacity calculations are almost 100 years old and it seems that their revision is an unfinished story. %has not yet come to an end.
Extensive opacity calculations that included for the first time bound-bound absorption started in Los Alamos
\citep{CoxStew1962,Cox1965} and were  known as the Los Alamos Opacity Library (LAOL) \citep{Hubner1977}.
For many years, these data were widely used but some disagreements were waiting to be explained, e.g., problems
with the standard solar model, an unknown mechanism of pulsations in B-type stars, %main-sequence
too large period ratios in classical Cepheids models. In the early nineties stellar opacities were recalculated
by two independent teams: OPAL \citep{Iglesias1992,Rogers1992} and OP \citep{Seaton1993,Seaton1994}.
The most spectacular result was the finding of a local maximum caused  by a huge number of transition lines of iron group elements.
This maximum occurs at temperature of about 200\,000 K and is called the $Z$-bump.

The discovery of the Z-bump was a big step forward in stellar physics, however there are still some uncertainties
and many indications that something is still missing and/or has not been correctly included \citep{Blancard2016}.
The first example is a disagreement between the standard solar model and the helioseismic and neutrino-flux predictions
\citep[e.g.,][]{Turck-Chieze2004,Guzik2008,Christensen-Dalsgaard2009} that arose after the revision of solar chemical abundances \citep{Asplund2005,Asplund2009}.
The laboratory measurements at physical conditions similar to the boundary of the solar convection zone
have indicated that the Rosseland mean opacities of iron predicted by all codes are underestimated by 30 to 400 \%
\citep{Bailey2015,Pradhan2018,Zhao2018}. The $30–45$\% underestimate of iron opacity at stellar interior temperatures
was also measured by \citet{Nagayama2019}.
%However a simple increase of opacity is not sufficient to solve problems in the solar modelling, e.g., \cite{Iglesias2017}.
Another example is a presence of high-order gravity modes in $\beta$ Cephei and $\delta$ Scuti stars, that are not excited
in standard-opacity models \citep[e.g.,][]{Pamyatnykh2004,Balona2014}. Increasing  the mean opacity at temperature of about 290\,000\,K,
where nickel has its maximum contribution to the Z-bump,  helped to excite g\,modes in $\beta$ Cep models \citep{Salmon2012,JDD2017}.
In the case of $\delta$ Scuti stars, an increase of  opacity at $T = 115\,000$\,K allowed to make g\,modes unstable \citep{Balona2015}.
This new opacity bump at $T = 115\,000$\,K was indeed identified by \citet{Cugier2012,Cugier2014}
in the Rosseland mean opacities taken from the  model atmospheres of \citet{Castelli2003}.

Seismic models, that is,  models that reproduce the observed frequencies of the identified pulsational modes, 
are also sensitive to the adopted opacity tables \citep[e.g.,][]{JDD2017,JDD2020,JDD2021,JDD2022}. 
Here, we present this effect for the double-mode radially pulsating $\delta$ Sct stars.
These pulsators are of special interest because the period ratio of  two consecutive radial modes  (usually the fundamental and first overtone)
takes the value in a very small range. On the other hand, we get  the period ratio from space observations with an accuracy of up
to six decimal places.  We performed extensive seismic studies, based on the Monte Carlo simulations, for the four  high-amplitude
 $\delta$ Sct (HADS) stars:   BP Peg, AE UMa, RV Ari and SX Phe. The first three stars belong to Population I and the last one, SX Phe,
belongs to Population II. One often talks about a separate group of pulsating variables  with SX Phe as their prototype.

The reminder of the paper is organized as follows. In Section\,2,  we explain the motivation for our studies.
Section\,3 contains the results of seismic modelling for the four HADS stars.
The last Section is the summary. Details of seismic modelling with the Bayesian analysis based on the Monte Carlo simulations
are given in Appendix.

\section{Motivation} \label{sec:motiv}

Three opacity databases  are commonly used in stellar evolution computations:
OPAL \citep{Iglesias1996}, OP \citep{Seaton1996, Seaton2005} and OPLIB \citep{Colgan2015,Colgan2016}.
There are some subtle differences between these data that result from the adopted physics and methods
of computations. However, they have, in general,  a minor effect  on evolutionary tracks. This is shown in the left panel of Fig.\,1,
where we plotted the tracks  in the Hertzsprung-Russell (HR) diagram, computed for a mass $M=1.8$\,M$_\odot$ with the OPAL, OP
and OPLIB opacities. %in the phase of main sequence and early post-main main sequnece.
For these tracks, we adopted a typical initial hydrogen abundance by mass $X_0=0.70$ and metallicity by mass $Z=0.014$.

Nevertheless, we found that opacity data has a very significant  effect on a frequency ratio of consecutive
radial  pulsational modes.  In the right panel of Fig.\,1, we show  the corresponding evolution of a frequency ratio of the fundamental
and first overtone radial modes. % as a function  of the effective temperature. % for a star with the mass $M=1.8$\,M$_\odot$.
Grey dots represent the models with the same values of the effective temperature  $T_{\rm eff}= 7030$\,K
and luminosity $\log L/$L$_\odot =1.2$. We selected these models to present the pure effect of opacity data
not blurred by the effect of $T_{\rm eff}$ or $\log L/$L$_\odot$.

Evolutionary computations were performed using  the Warsaw-New Jersey code, \citep[e.g.,][]{Pamyatnykh1999}.
The code takes into account the mean effect of the centrifugal force, assuming solid-body rotation
and constant global angular momentum during evolution. All the stars analysed in this paper are slow rotators,
which is one of the main properties of HADS stars. Therefore, the effect of differential rotation will be negligible.

The treatment of convection in the stellar envelope relies on the standard mixing-length theory (MLT) and the efficiency
of convection is described by the mixing length parameter $\alpha_{\rm MLT}$.
At lower temperature range, i.e., for $\log T<3.95$,  opacity data from \citet{Ferguson2005} were used.
The solar chemical mixture was adopted from \citet{Asplund2009}
and the OPAL2005 equation of state was used \citep{Rogers1996,Rogers2002}.
The evolutionary tracks in Fig.\,1 were computed at zero-rotation and for $\alpha_{\rm MLT}=0.5$, to focus only on the opacity effect.

Stellar pulsations were computed using linear non-adiabatic code of \citet{Dziembowski1977}.
This code adopts the frozen convection approximation, i.e., the convective flux does not change during the pulsations,
which is a reasonable approach if convection is not very efficient in the envelope.
The effects of rotation on pulsational frequencies are taken into account up to the second order in the framework of perturbation theory.
In the case of radial modes, the second order effects of rotation reduces to the factor  $4\nu_{\rm rot}^2 / 3\nu_{\rm puls}$ 
independently of the radial order $n$ \citep[e.g.,][]{Simon1969,Kjeldsen1998}, where $\nu_{\rm rot}$ is the frequency of rotation. 
At a rotation rate of 18\,km$\cdot$s$^{-1}$, 
for the dominant frequencies of the studied stars, this factor amounts to about $0.004\,\rm d^{-1}$  in the case of  SX\,Phe and about $0.002\,\rm d^{-1}$ in the case of other three HADS stars. For higher frequencies these values will be smaller.
\begin{figure*}
	\centering
	\includegraphics[clip,width=0.49\linewidth,height=64mm]{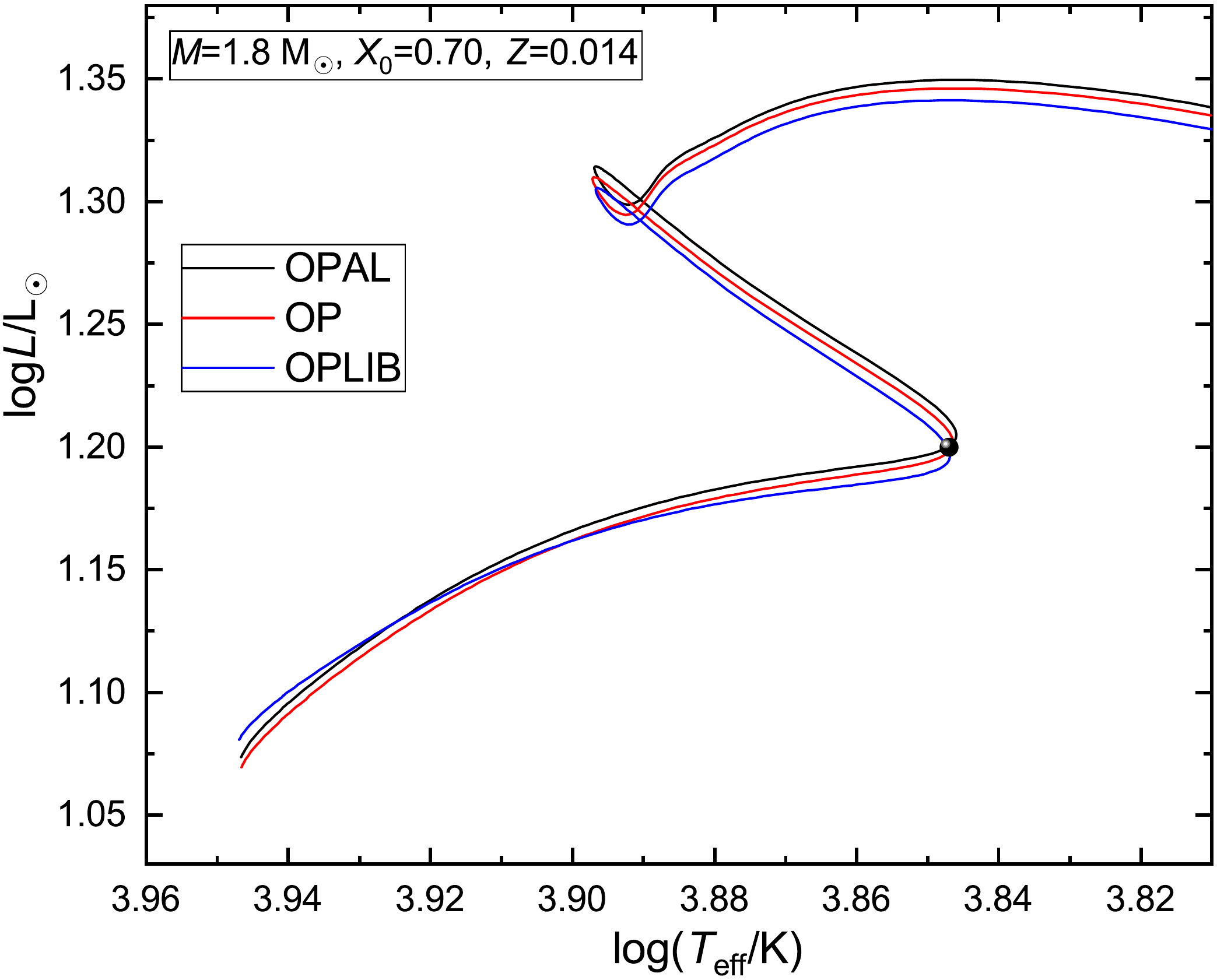}
	\includegraphics[clip,width=0.49\linewidth,height=64mm]{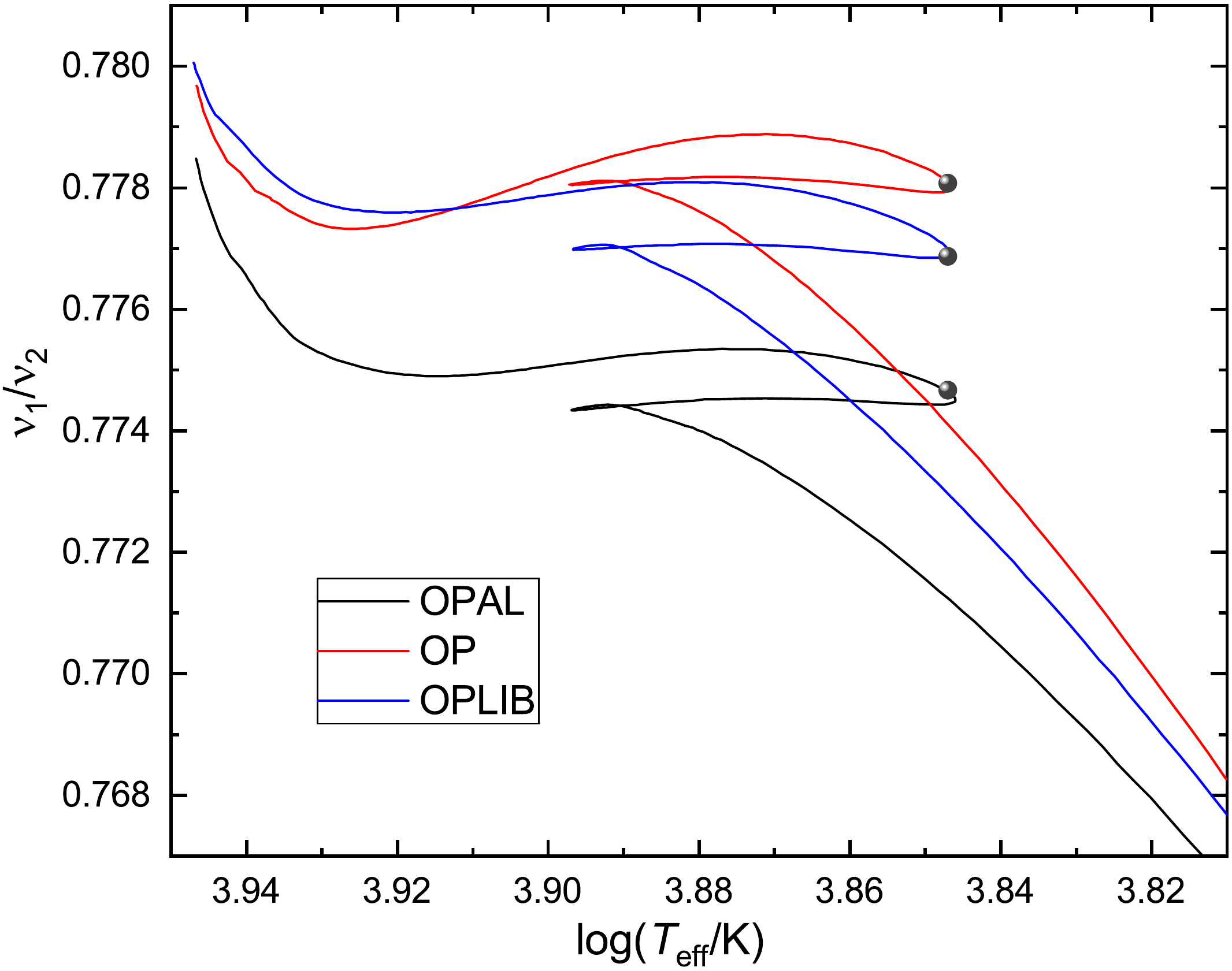}
	\caption{Left panel: evolutionary tracks of a 1.8\,M$_{\odot}$ star computed with the OPAL, OP and OPLIB opacity tables.
		Right panel: the evolution of a frequency ratio of the fundamental and first overtone radial modes.
		Grey dots represent models with the effective temperature, $T_{\rm eff} = 7020$\,K,  and luminosity, $\log L/$L$_{\odot}=1.2$,
		marked with a dot on the left panel.}
\end{figure*}
\begin{figure*}
	\centering
	\includegraphics[clip,width=0.49\linewidth,height=64mm]{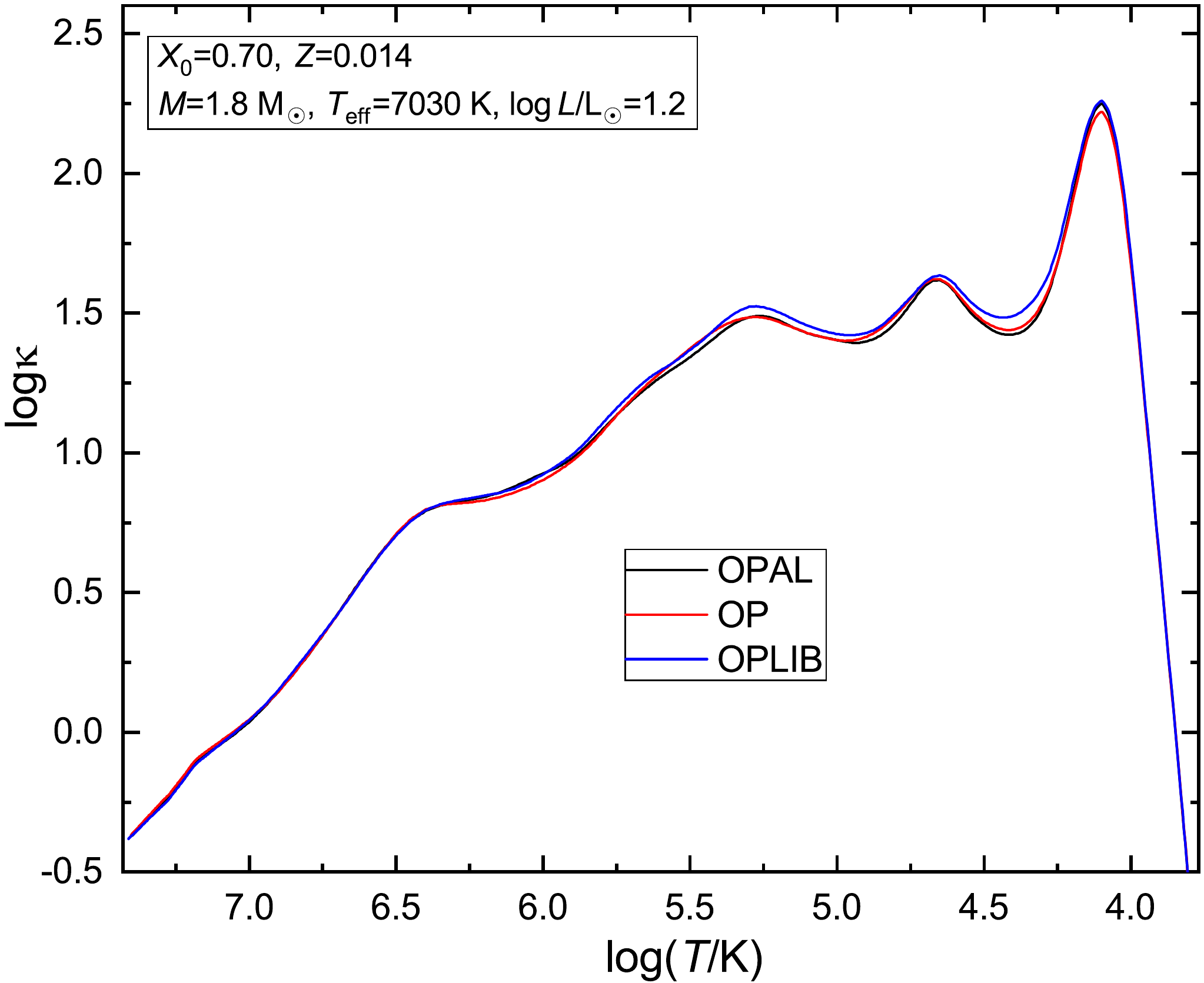}
	\includegraphics[clip,width=0.49\linewidth,height=64mm]{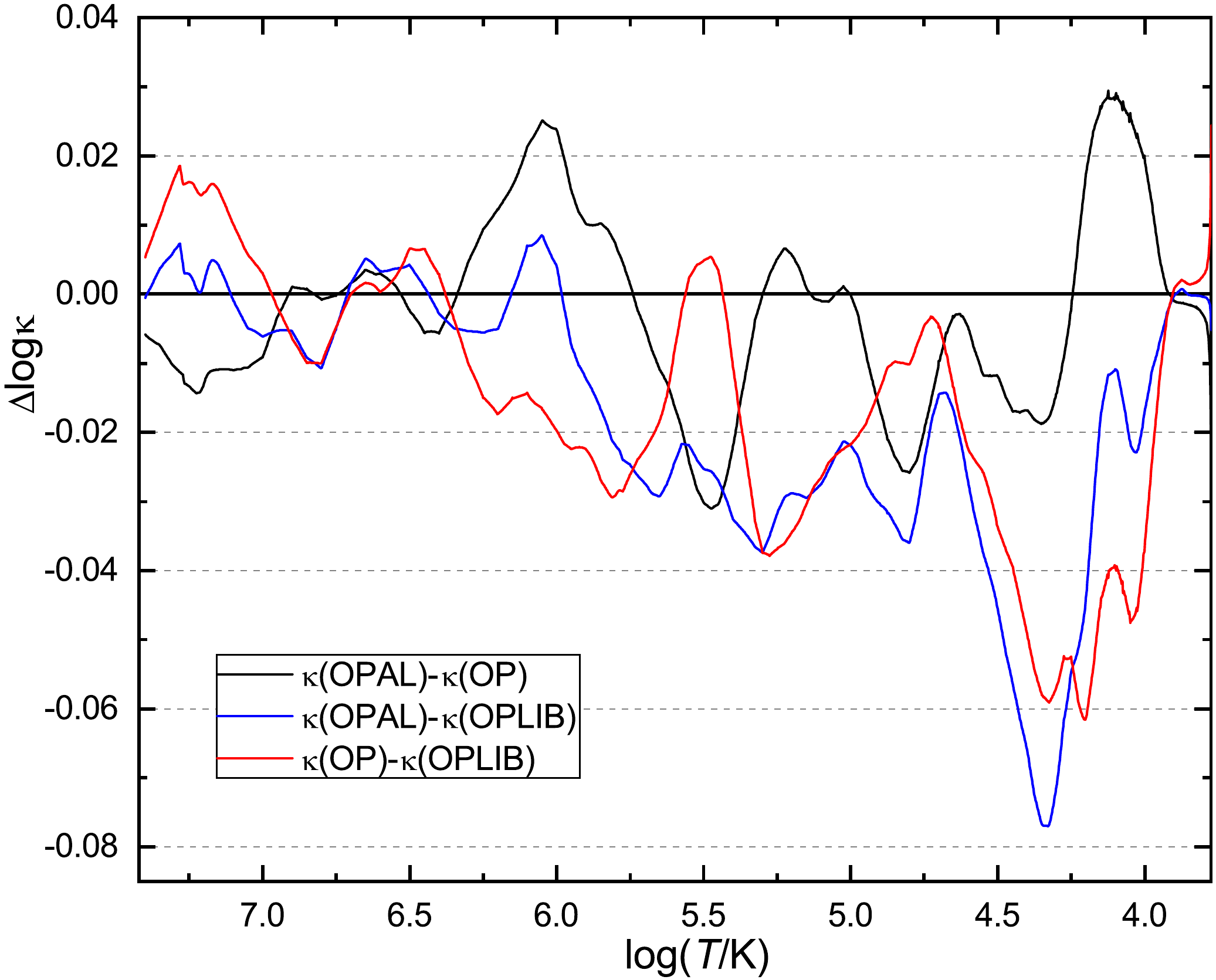}
	\caption{Left panel: the run of the Rosseland mean opacity from OPAL, OP and OPLIB data inside the model
		with a mass 1.8\,M$_{\odot}$, effective temperatures $T_{\rm eff} = 7020$\,K and luminosity $\log L/$L$_{\odot}=1.2$.
		%Data from different sources are marked with different colours.
		Right panel: the corresponding differences in $\log\kappa$.}
\end{figure*}

For the models marked with grey dots in Fig.\,1, we plotted  in Fig.\,2  the run of the Rosseland mean opacity.
As mentioned, all three models have the same effective temperature $T_{\rm eff}= 7030$\,K and luminosity $\log L/$L$_\odot=1.2$.
As one can see the overall run of $\log\kappa(T)$ from the three databases is very similar.
However, by drawing the differences, it is possible to locate certain depths where three opacity profiles deviate from each other.
This is shown in the right panel of Fig.\,2.  For example, there are significant differences between OPAL, OP and OPLIB around
temperature corresponding to the hydrogen  and first helium ionization  ($\log T\approx 4.0 - 4.45$).  On the other hand,
all three $\kappa$'s are very close in the second helium ionization zone ($\log T\approx 4.67$), where main driving
of pulsations for $\delta$ Sct stars occurs. One can also see that OPAL and OP are very close to each other near the $Z$-bump,
i.e., near $\log T=5.2-5.3$. Instead, the data from OP and OPLIB are very similar near $\log T\approx 5.46$, where nickel has
its maximum contribution to the opacity around $Z$-bump.

All these small differences in opacity tables (of a few precent) cause huge differences in the frequency ratio,
already in the third  decimal place, while modern observations give accuracy down to six decimal places.
In the next section, we will study how these differences affect seismic models of the realistic stars.

\section{Seismic models of the four high-amplitude $\delta$ Scuti pulsators} \label{sec:seismic}
\begin{table*}
	\centering
	\caption{The main parameters from observations of the four studied $\delta$ Sct stars, i.e., the range of effective temperature,
		luminosity, metallicity and the projected rotational velocity.}
	\begin{tabular}{c c c l r c c } % four columns, alignment for each
		\hline
		star      & Pop.         &    $T_{\rm eff}$  &   $\log L/$L$_\odot$  &  [m/H] & $V_{\rm rot}\sin i$   \\
		&           &  [K ]     &       &   [dex] &   [km$\cdot$s$^{-1}$]\\
		\hline
		BP Peg    &  I &     6800$-$8100 &  1.247(120)  & $0.2$  &  $\sim$18 \\
		AE UMa  &  I &    7100$-$8200 &  1.091(90)     & $-0.3$ &  $<$10  \\
		RV Ari    &  I &    7000$-$8200 &   1.103(26)    & $0.1$   &  $\sim$18 \\
		SX Phe   &  II &   7000$-$8600 &   0.899(17)    & $-1.0$ &  18(2)  \\
		\hline
	\end{tabular}
\end{table*}

To study the effect of opacity, we selected four high-amplitude $\delta$ Sct stars: BP Pegasi, AE Ursa Majoris, RV Arietis
and SX Phoenicis.  These are relatively simple objects because: 1) they pulsate in the two radial modes: fundamental and first overtone,
2)  the linear theory of pulsations is still applicable, because  even in the case of  classical Cepheids, which have the light amplitudes
about five times larger, nonlinear period ratios differ from  linear values only by several tenths of a per cent  \citep[e.g.,][]{Kollath2001}.
3)  they are very slow rotators, 4) their effective temperatures are not low enough
for convection in their envelopes to be efficient, 5) mass loss can be neglected, and 6) there are no observational evidences for  anomalous surface abundances, so in the first approximation the effect of diffusive settling or radiative levitation can be safely neglected.

Some results for SX\,Phe and BP\,Peg were already presented in \citet{JDD2020} and \citet{JDD2022}, respectively.
Here, we included the Monte Carlo simulations for SX\,Phe and increased the number of simulations for BP Peg.

The basic parameters of the studied stars are given in Table\,1. The range of the effective temperature were gathered from the
literature  and luminosity were derived from the Gaia DR3 data \citep{Gaia2022} in this paper.
This $T_{\rm eff}$ range also includes the changes due to pulsations.
The last two columns contain the metallicity [m/H] and projected rotational velocities $V_{\rm rot}\sin i$.
These parameters were adopted from \citet{Rodriguez1992} for BP Peg, AE UMa and RV Ari, and from \citet{Antoci2019} for SX\,Phe.

In Table\,2, we list the observed frequencies of the two radial modes and  the corresponding amplitudes.
In the case of BP Peg, we performed the frequency analysis using the ASAS photometry \citep{Pojmanski1997}. For the other three stars
TESS observations were available and we used them to determine the frequencies. We know that in each case these two frequencies
are radial modes, based on the period ratio and  on the  independent mode identification from multi-colour photometry \citep{JDD2020,JDD2022,JDD2023}.
\begin{table*}
	\centering
	\caption{Frequencies of the fundamental ($\nu_1$) and first overtone ($\nu_2$) radial modes of the four studied HADS stars.
		The frequencies of BP Peg were derived from the ASAS data and their amplitudes are given in [mmag].
		For the other three stars the frequencies were obtained from the TESS light curves and the amplitudes are expressed in [ppt]. }
	\begin{tabular}{c|ll|ll|cc}
%		\hline
%		&  \multicolumn{2}{c|}{fundamental }   &   \multicolumn{2}{c|}{first overtone}  & \\
%		&   \multicolumn{2}{c|}{radial mode}  &   \multicolumn{2}{c|}{radial mode}    &     \\
		\hline
		star   &   ~~~~~~$\nu_1$   &   Ampl.   &  ~~~~~~$\nu_2$  &   Ampl.  & $\nu_1/\nu_2$ \\
	           	&  ~~~~[d$^{-1}$] &           &  ~~~~[d$^{-1}$]  &              &     \\
		\hline
		BP Peg   &     ~~9.128797(4)      &      208(4)         &     11.83315(3)        &   34(4)         &  0.77146 \\%		& & & & & \\
		AE UMa &   11.625598(2)     & 131.65(2)       &     15.031250(5) &   30.70(2)      &   0.77343    \\%		& & & & & \\
		RV Ari   &   10.737888(55)   &   128.84(3)          &      13.899137(121)    &  39.62(2)     &    0.77256       \\%		& & & & & \\
		SX Phe  &    18.193566(3)   &    133.64(1)     &      23.379306(7)      &      32.92(10)   &  0.77819   \\
		\hline
	\end{tabular}
\end{table*}

Having the two radial modes, fundamental and first overtone, we constructed seismic models that fit the observed frequencies
within the errors.  %Besides, we fitted the effective temperature and luminosity as well as the photometric Str\"omgren amplitudes and phases for the dominant mode.
Besides, we fitted also the non-adiabatic parameter $f$ for the dominant frequency. This parameter is a relative amplitude
of radiative flux variations at the level of the photosphere. The theoretical value of $f$  for a given pulsational mode is derived
in the framework of non-adiabatic theory of stellar pulsations and it is complex because there is a phase shift between the flux
and radius variations. In the case of $\delta$ Sct models, the parameter $f$ is very sensitive to the adopted value
of $\alpha_{\rm MLT}$ \citep[e.g.,][]{JDD2003}. Therefore,  from matching the theoretical and empirical values of $f$ it is possible
to get reliable constraints on the efficiency of convection in sub-photospheric layers.
The empirical values of $f$ are derived from the photometric amplitudes and phases in at least three passbands, see \citet{JDD2020,JDD2021,JDD2022} for recent results.

Our extensive, complex seismic modelling was made using the Bayesian analysis based on the Monte Carlo simulations.
As a result,  we obtained constraints on the mass $M$, metallicity $Z$, initial hydrogen abundance $X_0$,
rotational velocity $V_{\rm rot}$ as well as on the mixing length parameter ($\alpha_{\rm MLT}$).
In the case of  BP Peg, AE UMa and RV Ari (Population I stars),  we got $\alpha_{\rm MLT}\approx 0.6,~0.4,~0.5$, respectively,
and in the case of SX Phe (the Population II star), we obtained $\alpha_{\rm MLT}\approx 1.0$. Details of our computations
and detailed results are given in the Appendix.

Results for $X_0=0.70$ and some range of $Z$ are presented in Fig.\,3 in the HR diagram. As one can see, in the case of each star, only
the OPAL seismic models have effective temperatures and luminosities consistent with the observational determinations.
The OP and OPLIB seismic models have far too low values of $\log T_{\rm eff}$ and $\log L/$L$_\odot$.
In the case of the Population I stars, the OP models have parameters even lower than the OPLIB models.
In the case of SX Phe, the OP and OPLIB seismic models are in the same position in the HR diagram.
All depicted models, for each set of opacity data, are in the post-main sequence phase of evolution.
The vast majority of them burn hydrogen in the shell surrounding the core and just one or two models are in the phase 
of an overall contraction. All seismic models in the main sequence phase of evolution had much too low $T_{\rm eff}$ 
and $\log L/$L$_\odot$ in the case of each opacity tables. Both radial modes, fundamental and first overtone,
are unstable.

\begin{figure*}
	\centering
	\includegraphics[clip,width=0.49\linewidth,height=69mm]{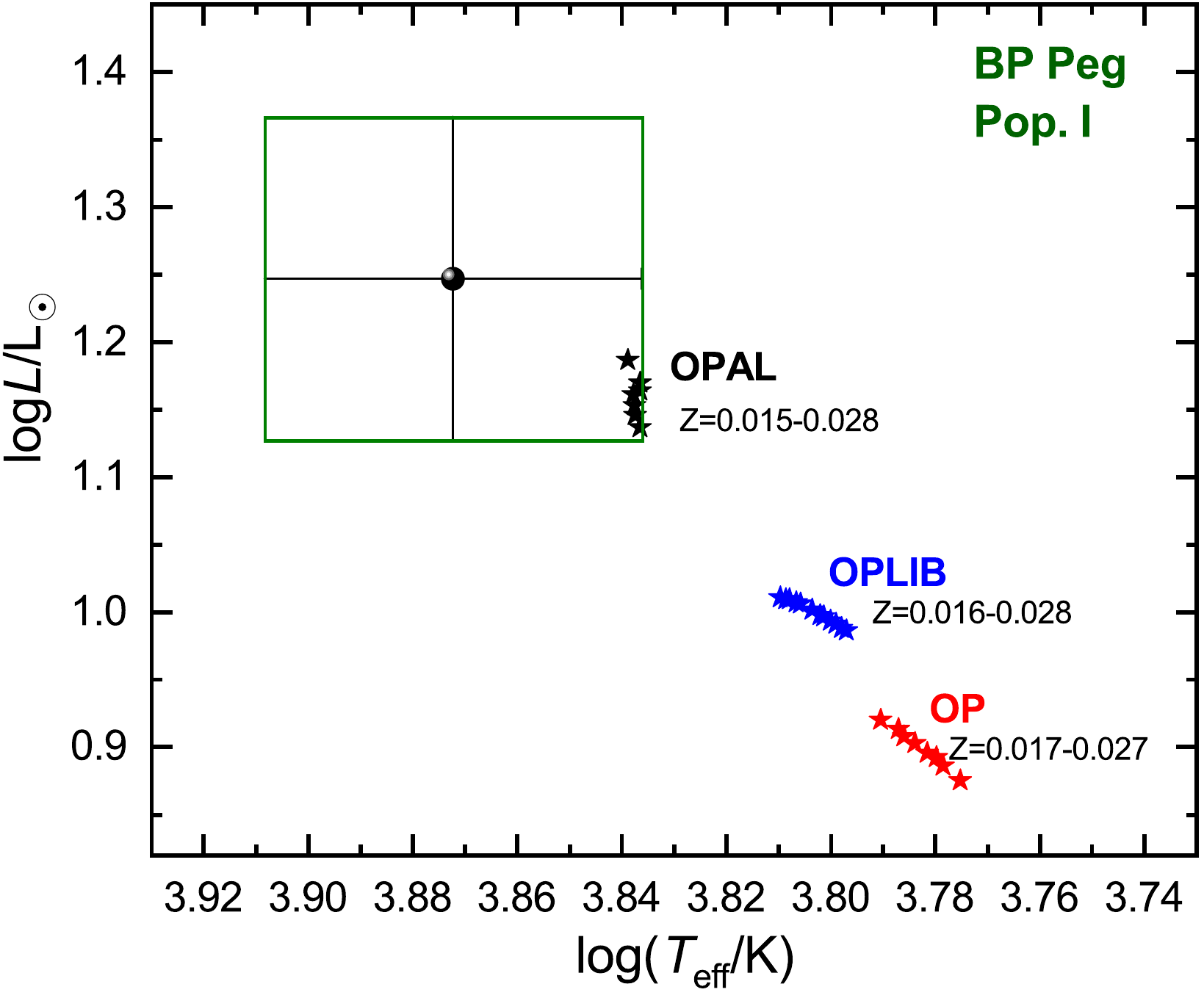}
	\includegraphics[clip,width=0.49\linewidth,height=69mm]{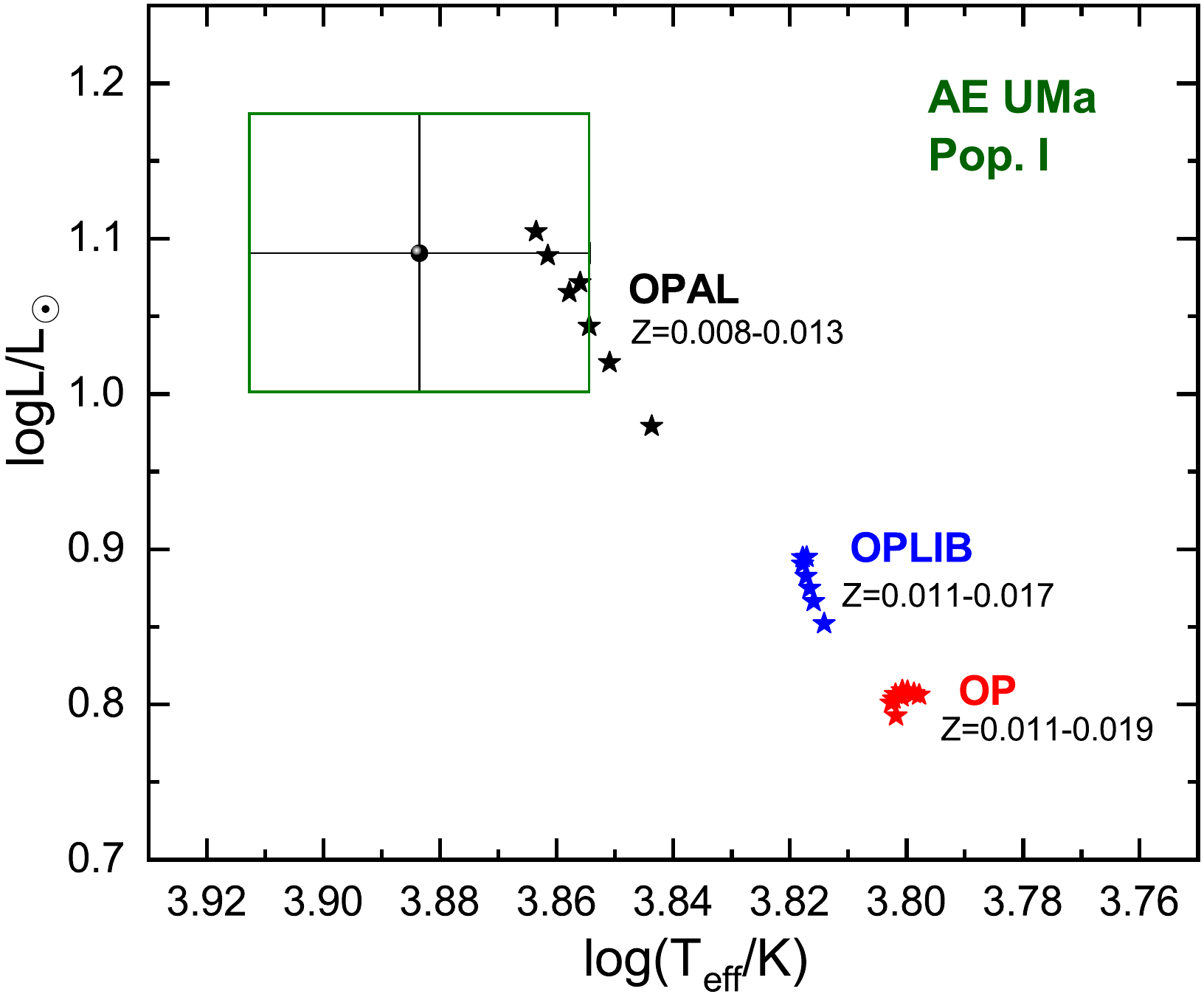}
	\includegraphics[clip,width=0.49\linewidth,height=69mm]{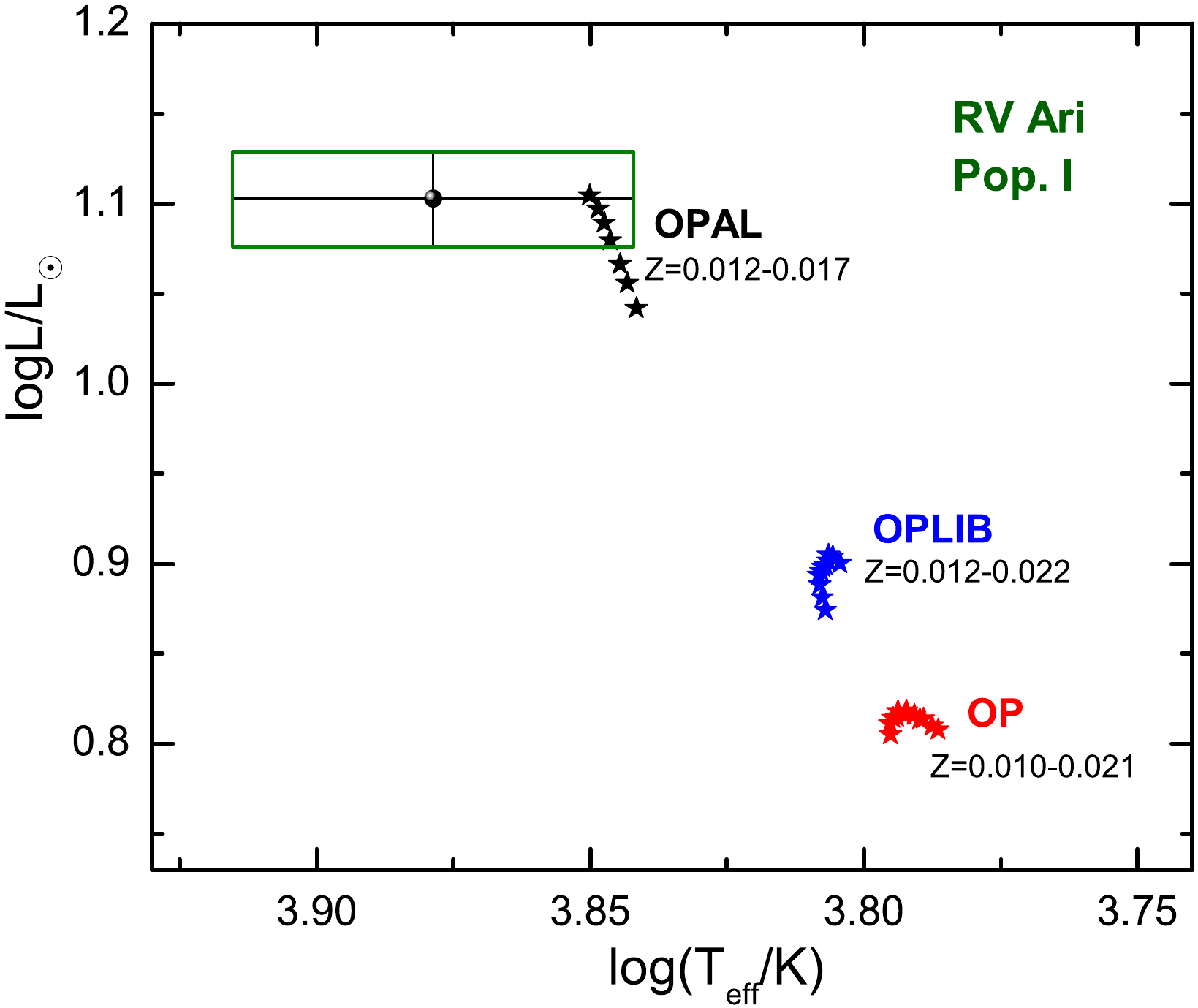}
	\includegraphics[clip,width=0.49\linewidth,height=68mm]{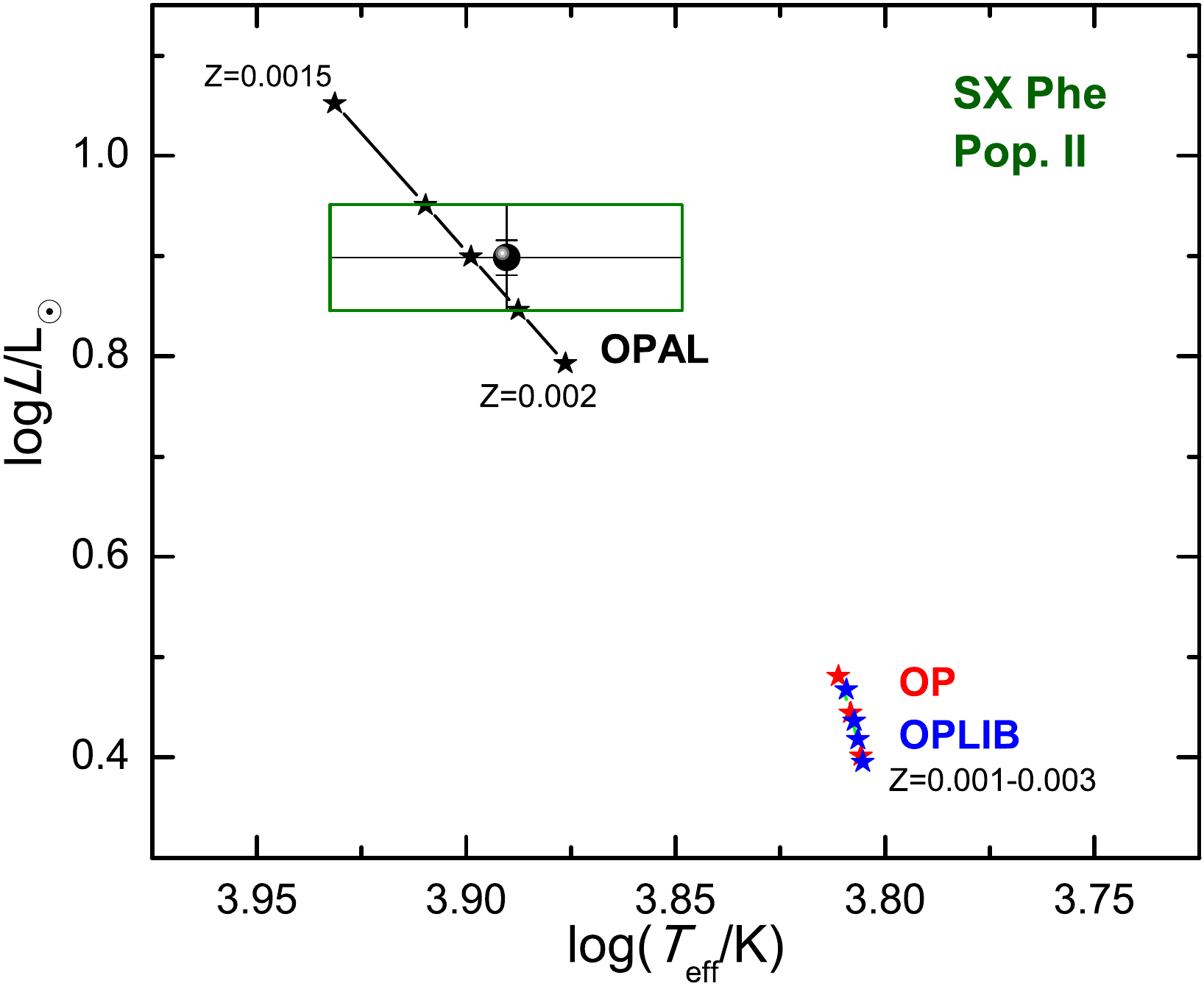}
	\caption{Seismic models of BP Peg, AE UMa, RV Ari and SX Phe constructed with the three opacity data: OPAL, OP and OPLIB.
		The initial hydrogen abundance by mass was $X_0=0.70$ and the range of metallicity $Z$ is given in each case. }
\end{figure*}

\section{Summary} \label{sec:summary}

 We presented the results of extensive seismic modelling for the four high-amplitude $\delta$ Sct stars that
pulsate in two radial modes, i.e, BP Peg, AE UMa, RV Ari and SX Phe. To this aim we used the Bayesian analysis
based on the Monte Carlo simulations. Three opacity tables were adopted: OPAL, OP and OPLIB.
Besides frequencies, we fitted also the parameter $f$ for the dominant frequency  in order to constrain, in particular,
the mixing length parameter $\alpha_{\rm MLT}$ in the envelope.
We obtained $\alpha_{\rm MLT}\approx 0.6, 0.4, 0.5$ for BP Peg AE UMa and RV Ari,  respectively,
and $\alpha_{\rm MLT}\approx 1.0$ for SX Phe.

In the case of each studied HADS, only the OPAL seismic models are close to or within the error box
on the HR diagram whereas the OP and OPLIB seismic models are far beyond.
This "seismic opacity" discrepancy is independent of the metallicity as it was obtained for stars
with different values of [m/H] and even for a Population II HADS SX Phe.
Thus, we have shown that there are systematic differences in seismic models of double-mode radial $\delta$ Sct pulsators,
computed with different opacity data.
%This is quite a surprising result and
However, the solution to this puzzle is rather beyond our ability.
Rather, this is one more message to atomic physicists that something is still missing in stellar opacities.
This is also a warning to those who model double-mode classical Cepheids or RR Lyr stars.
Such huge effect of opacity can occur also for these variables which are used as standard candles to measure distances.

\begin{acknowledgments}
The work was financially supported by the Polish National Science Centre grant 2018/29/B/ST9/02803.
Calculations have been partly carried out using resources provided by Wroclaw Centre for Networking and Supercomputing (http://www.wcss.pl), grant No. 265.
This work has made use of data from the European Space Agency (ESA) mission {\it Gaia} (https://www.cosmos.esa.int/gaia), 
processed by the {\it Gaia} Data Processing and Analysis Consortium (DPAC,
https://www.cosmos.esa.int/web/gaia/dpac/consortium).
Data collected by the TESS mission and ASAS project were used. Funding for the TESS mission is provided
by the NASA's Science Mission Directorate.
\end{acknowledgments}

\bibliography{JDD_apjL}{}
\bibliographystyle{aasjournal}

\appendix

\section{Asteroseismic modelling with Monte Carlo-based Bayesian analysis}
We computed extensive grids of seismic models for each star and employed Monte Carlo–based Bayesian analysis
to obtain  constraints on various parameters.
The analysis was based on the Gaussian likelihood function defined as \citep[e.g.,][]{2005A&A...436..127J, 2006A&A...458..609D, 2017MNRAS.467.1433R,Jiang2021}
\begin{equation}
{\cal L}(E|{\mathbf H})=\prod_{i=1}^n \frac1{\sqrt{2\pi\sigma_i^2}} \cdot
{\rm exp} \left( - \frac{  ({\cal O}_i-{\cal M}_i)^2}{2\sigma_i^2}  \right),
\end{equation}
where ${\mathbf H}$ is the hypothesis that represents adjustable model and theory parameters that in case of studied stars were:
mass $M$, initial hydrogen abundance $X_0$, metallicity $Z$, initial rotational velocity $V_{\rm rot,0}$,
convective overshooting parameter $\alpha_{\rm ov}$ and the mixing length parameter $\alpha_{\rm MLT}$.
The evidence $E$ represents the calculated observables ${\cal M}_i$, %e.g.,  the effective temperature $T_{\rm eff}$, luminosity $L/L_{\odot}$, pulsational frequencies,
that can be directly compared with the observed parameters ${\cal O}_i$ determined with the errors $\sigma_i$.

Here, we used the following observations: effective temperature $T_{\rm eff}$, luminosity $L/L_{\odot}$,
the frequencies of the two radial modes $\nu_1$ and $\nu_2$, and the non-adiabatic parameter $f$ for the dominant mode.
The second mode $\nu_2$ had too low amplitudes to make use of it. To derive the empirical values of $f$
we applied the method of   \citet{JDD2003}. To this aim, we used the amplitude and phases in the four Str\"omgren passbands
from \citet{Rodriguez1992} for BP Peg, AE UMa and RV Ari,  and from \citet{Rolland1991}  for SX Phe.
The method requires also the model atmospheres and here we relied on Vienna (NEMO) models \cite{Heiter2002} that include 
turbulent convection treatment from \cite{Canuto1996}.

Then, we made a huge number of simulations (from about 90\,000 to 160\,000 depending on the star) to maximize
the likelihood function given in Eq.\,(A1) in order to constrain the parameters for each star.
For each randomly selected set of parameters ($M,~X_0,~Z,~V_{\rm rot,0},~\alpha_{\rm ov}$ and $\alpha_{\rm MLT}$),
we calculated evolutionary and pulsational models.
%Then, we chose the models that reproduce all observables, i.e., the observed frequencies of the radial fundamental and first overtone, effective temperature and luminosity within the observational errors,  the photometric amplitudes and phases of the dominant mode $\nu_1$.

In the case of the initial hydrogen abundance $X_0$, we assumed a beta function $B(2,2)$ as a prior probability,
since we wanted to limit its value to the reasonable range, i.e., from 0.65 to 0.75 with $X_0=0.7$ as the most probable.
For other parameters we used uninformative priors, i.e., a uniform distribution. Moreover, already first simulations showed that 
overshooting for the convective core is ineffective because  in all runs the parameter $\alpha_{\rm ov}$ tended to zero very fast and did not change. This is because all models with $\alpha_{\rm ov}>0$ had much too low values of $\log L/$L$_\odot$.
Therefore, we set $\alpha_{\rm ov}=0.0$ in further computations.

\section{Constraints on parameters}
The most important result of our seismic modeling is that only with the OPAL opacity tables, we were able to match all observables for each star.
In the case of OP and OPLIB data, it was often difficult to find any model that reproduced both frequencies and the non-adiabatic $f$ 
for the dominant mode. Even if such models were found, they were far beyond the error box in the HR diagram. 
%Moreover, the value of the microturbulent velocity in the atmosphere at which the empirical value of $f$ was best reproduced was $\xi_t=4\,$km$\cdot$\,s$^{-1}$.

The results of our simulations with the OPAL data for all studied stars are presented as histograms.
The histograms were normalised to 1.0  by the number of all models, thus the numbers on the Y-axis times 100
are the percentage  of models with a given parameter range.
In Fig.\,4, there are shown histograms for masses of BP Peg, AE UMa, RV Ari and SX Phe.
The next Figures\,5$-$8, present histograms for $Z,~X_0$, the current value of rotation $V_{\rm rot}$ and for $\alpha_{\rm MLT}$.
The expected values of the parameters from these distributions  as well as of the initial rotation, effective temperature and  luminosity
are given in Table\,3.  The errors in parentheses are standard deviations. The current value of rotation $V_{\rm rot}$ corresponds
to the given seismic model.
In Table\,4, we give the median values for the determined parameters.  This statistic is more informative
for skewed distributions or distributions with outliers, which is the case for some parameters.
For example, in the case of  BP Peg, AE UMa and RV Ari, rather an upper limit may be given for the rotational velocity. 
One can see also the asymmetry for the metallicity $Z$ in the case of BP Peg,
AE UMa and RV Ari,  while the histogram of $Z$ for SX Phe shows quite a symmetric distribution.
On the other hand, the histograms for the mixing parameter $\alpha_{\rm MLT}$ are almost symmetric
for BP Peg , AE UMa and RV Ari, while in the case of SX Phe we have some skewness.
The errors in Table\,4 were calculated as the 0.84-quantile minus the median and the median minus the 0.16-quantile.
These quantiles correspond to estimates of values separated by one standard deviation
from the mean value in the case of a normal distribution.

In the case of each star, all presented models are in the post-main sequence phase of evolution, mostly during the hydrogen-shell burning
and very rare in the overall contraction. Main sequence models had always too low effective temperatures and luminosities.
Efficiency of convection in the envelopes of the Population I stars, parameterized by $\alpha_{\rm MLT}$,
is rather low and amounts to about 0.5. In the case of SX Phe, which is a lower-mass star and belongs to Population II,
convection is more efficient and $\alpha_{\rm MLT}\approx 1.0$.
\begin{figure*}
	\centering
	\includegraphics[clip,width=0.49\linewidth,height=63mm]{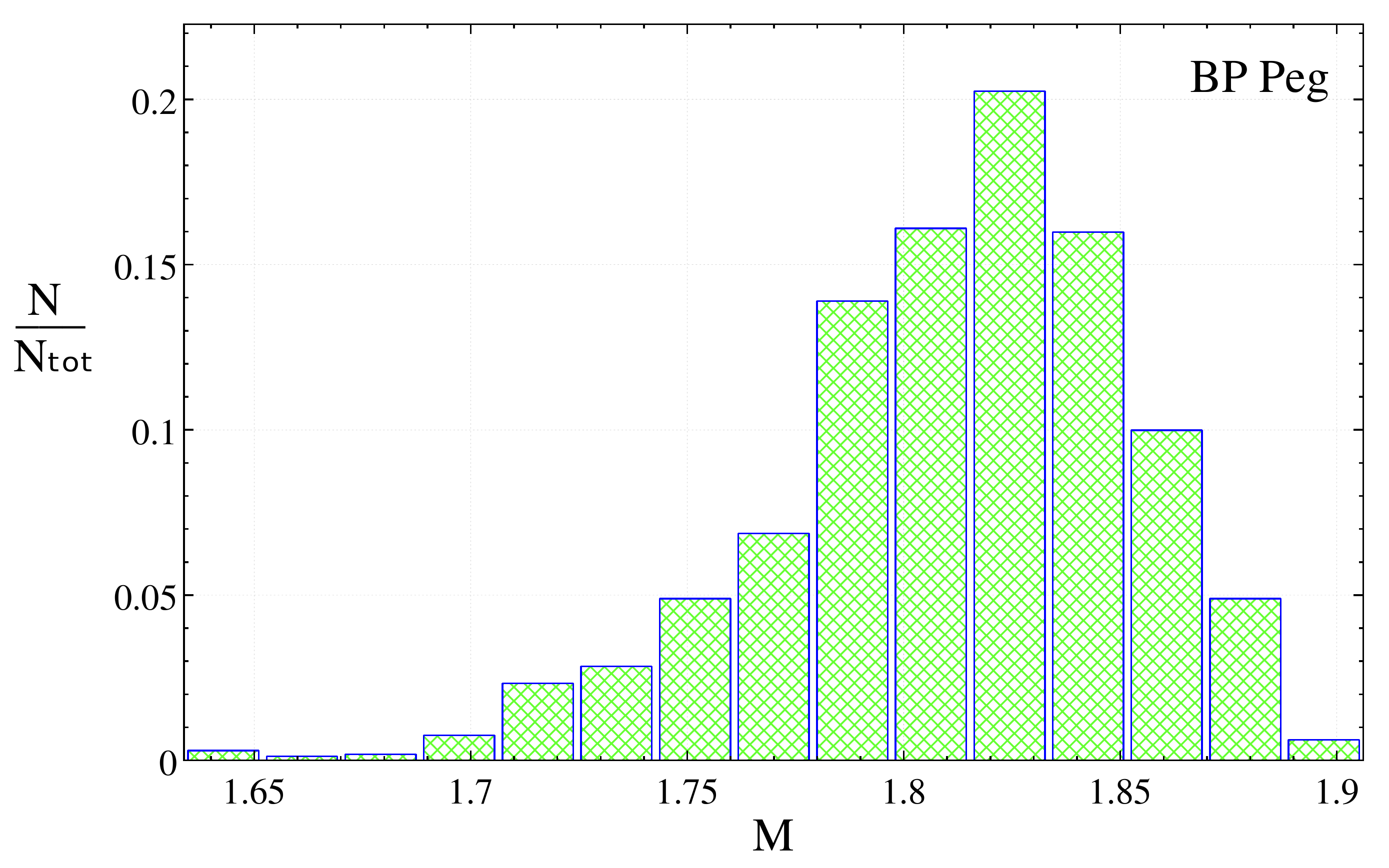}
	\includegraphics[clip,width=0.49\linewidth,height=63mm]{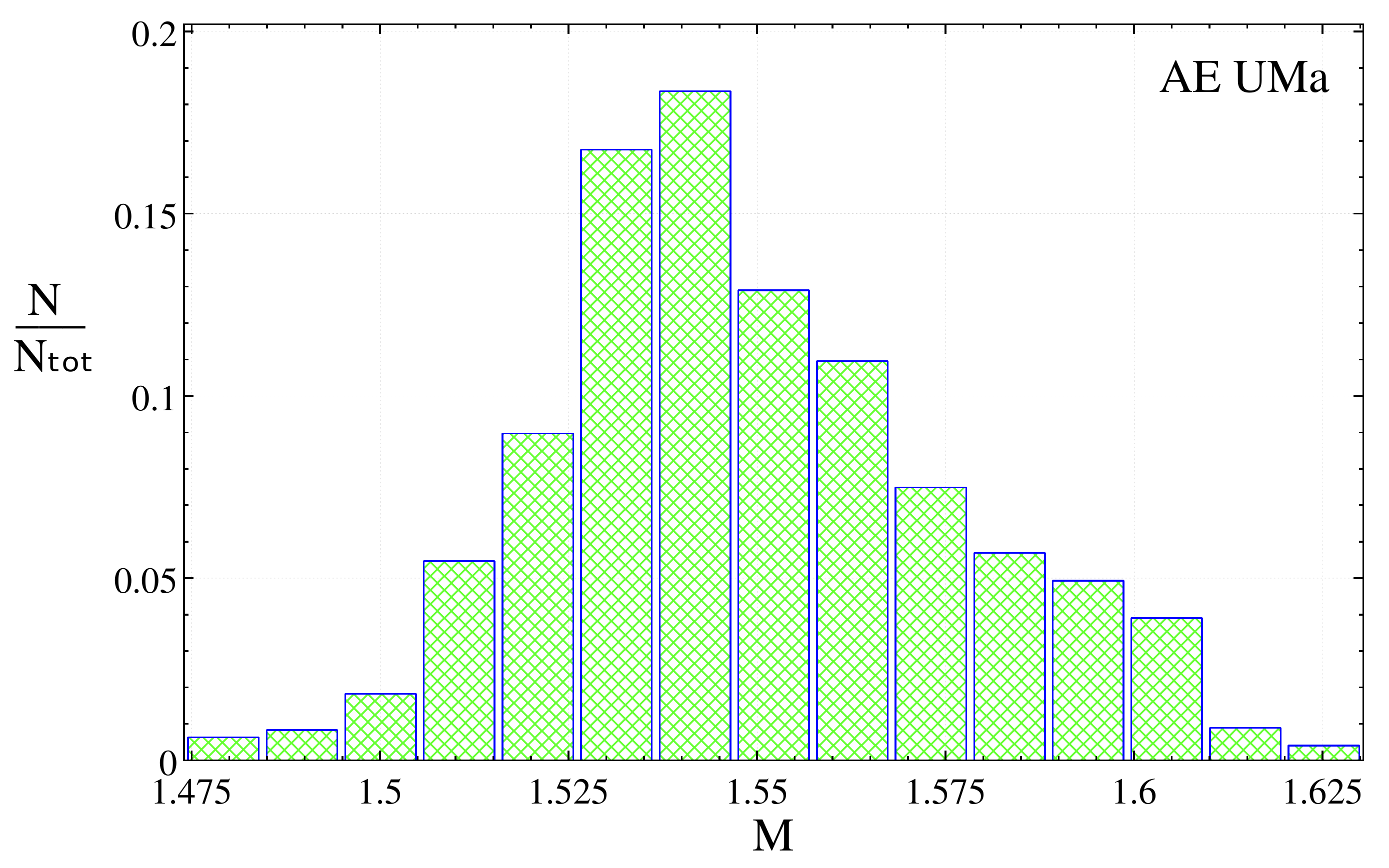}
	\includegraphics[clip,width=0.49\linewidth,height=63mm]{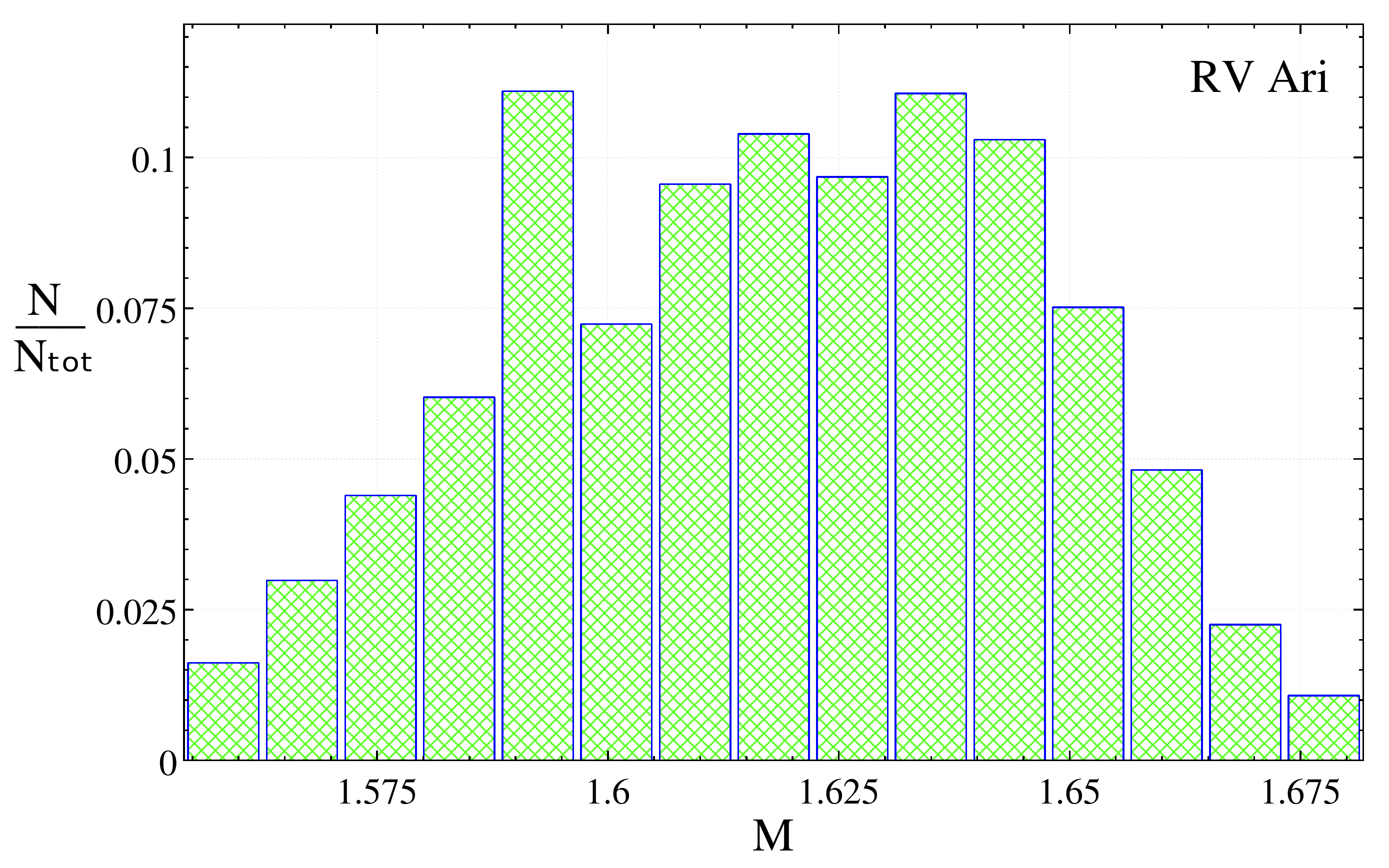}
	\includegraphics[clip,width=0.49\linewidth,height=63mm]{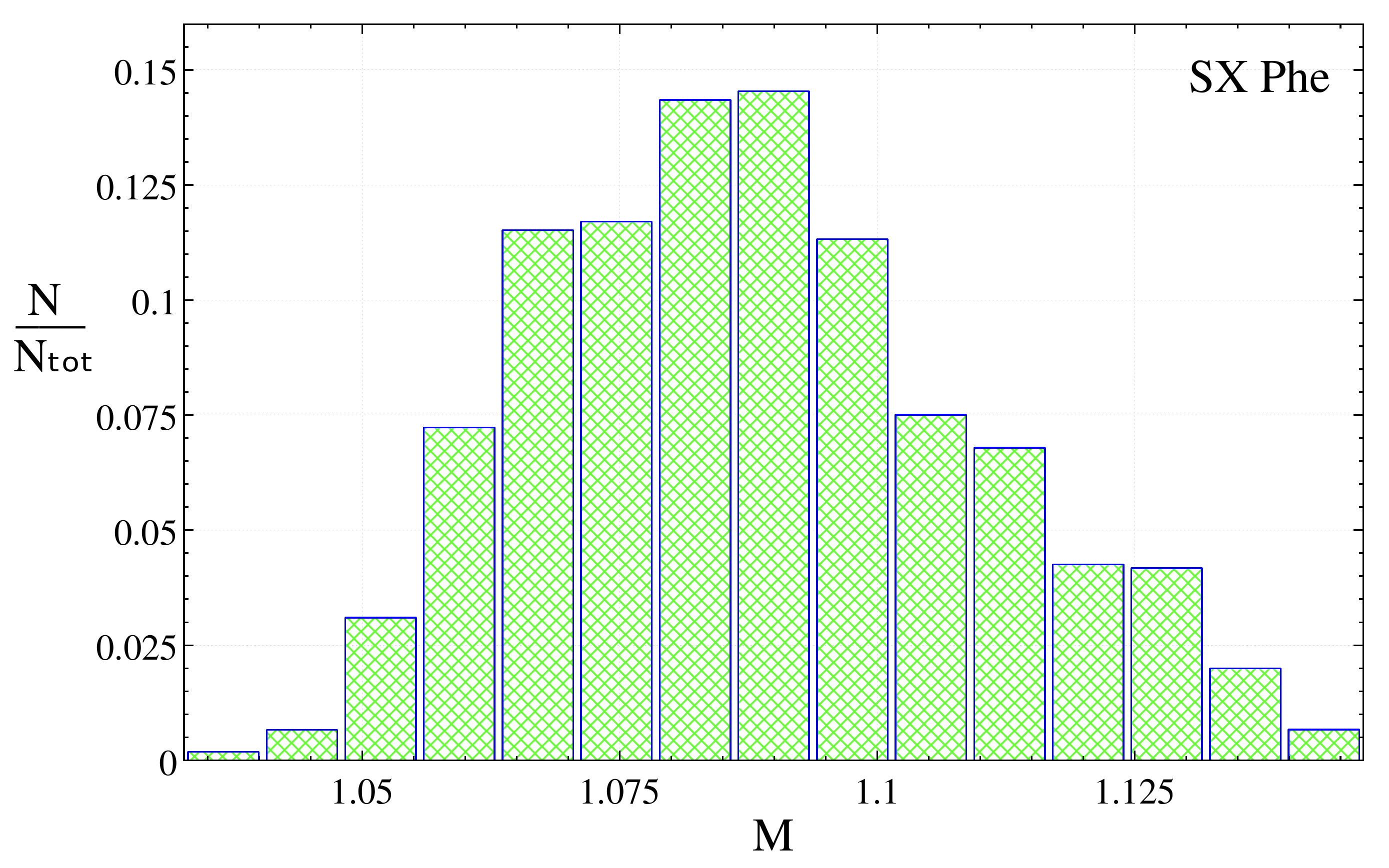}
	\caption{The normalized histograms for the mass  (in M$_\odot$) of  BP Peg, AE UMa, RV Ari and SX Phe.
All seismic models were computed with the OPAL opacities.}
\end{figure*}
\begin{figure*}
	\centering
	\includegraphics[clip,width=0.49\linewidth,height=63mm]{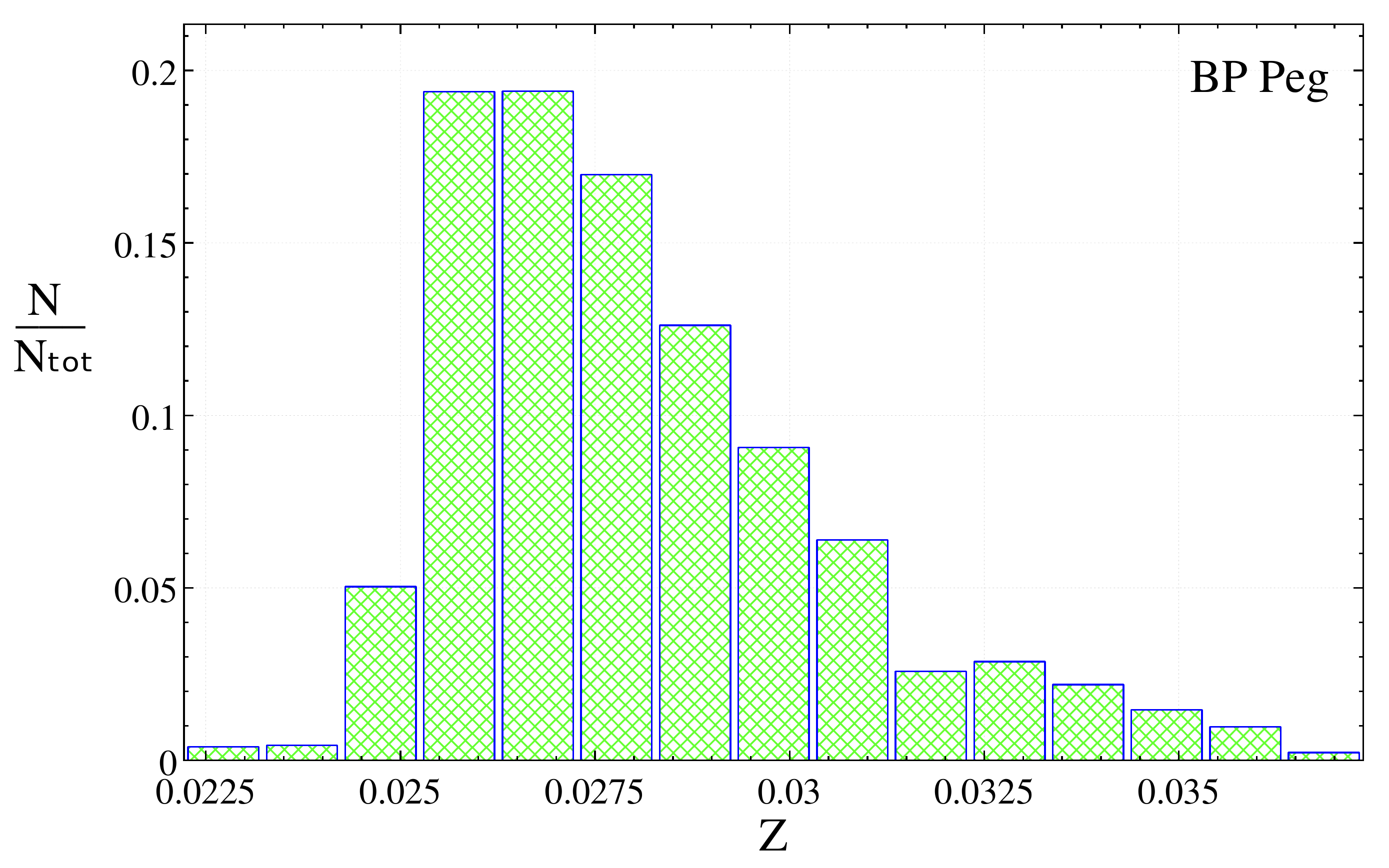}
	\includegraphics[clip,width=0.49\linewidth,height=63mm]{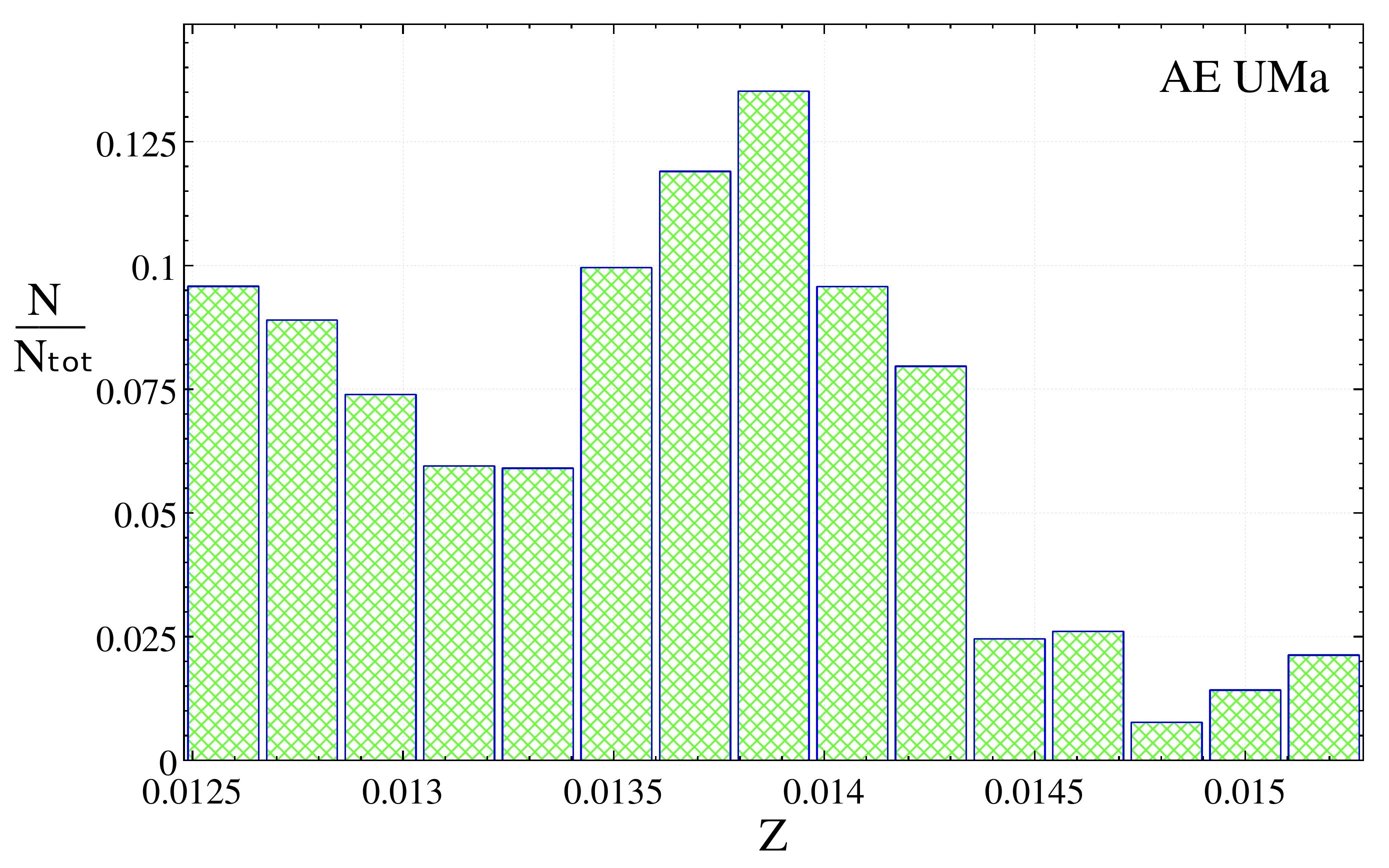}
	\includegraphics[clip,width=0.49\linewidth,height=63mm]{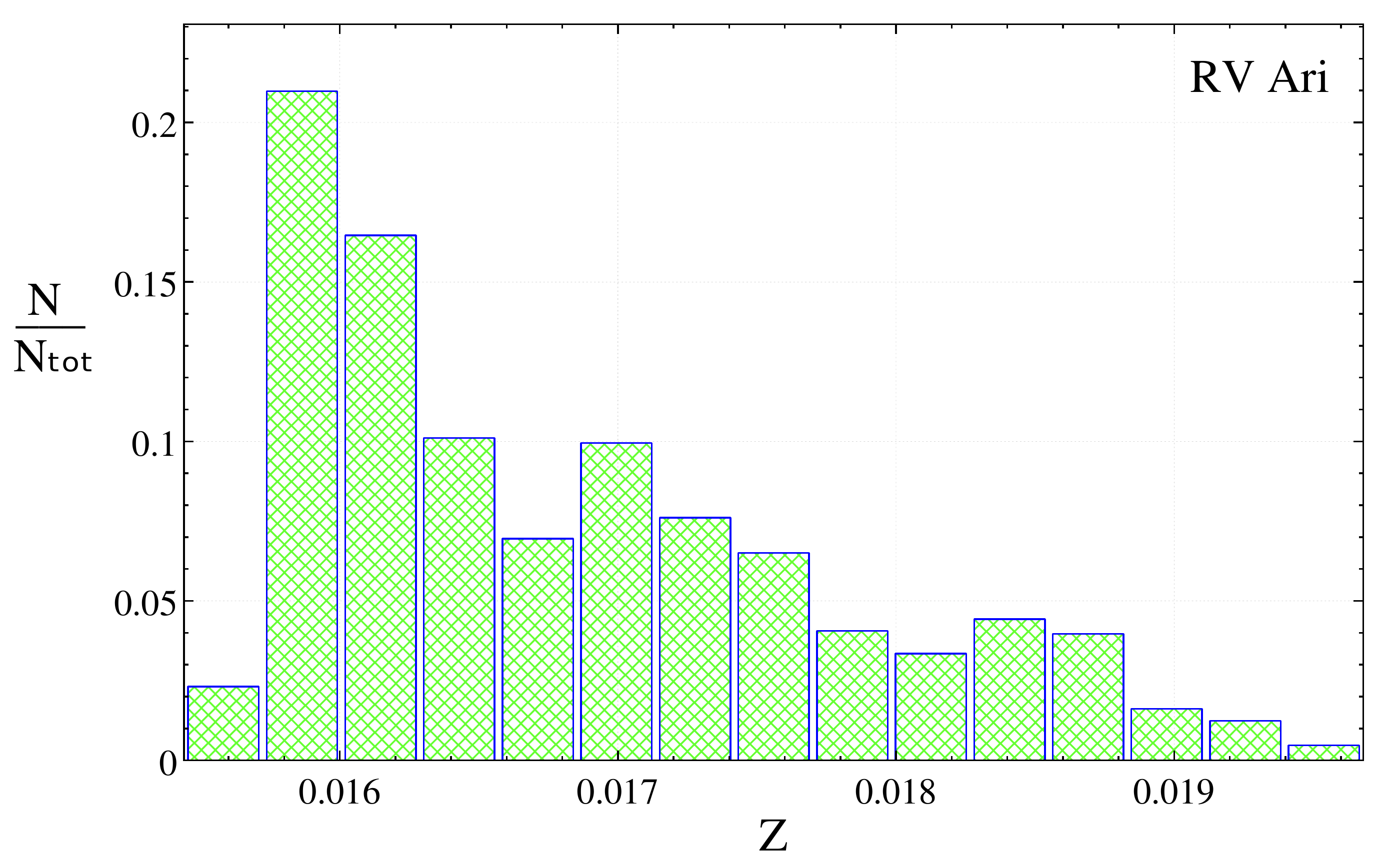}
	\includegraphics[clip,width=0.49\linewidth,height=63mm]{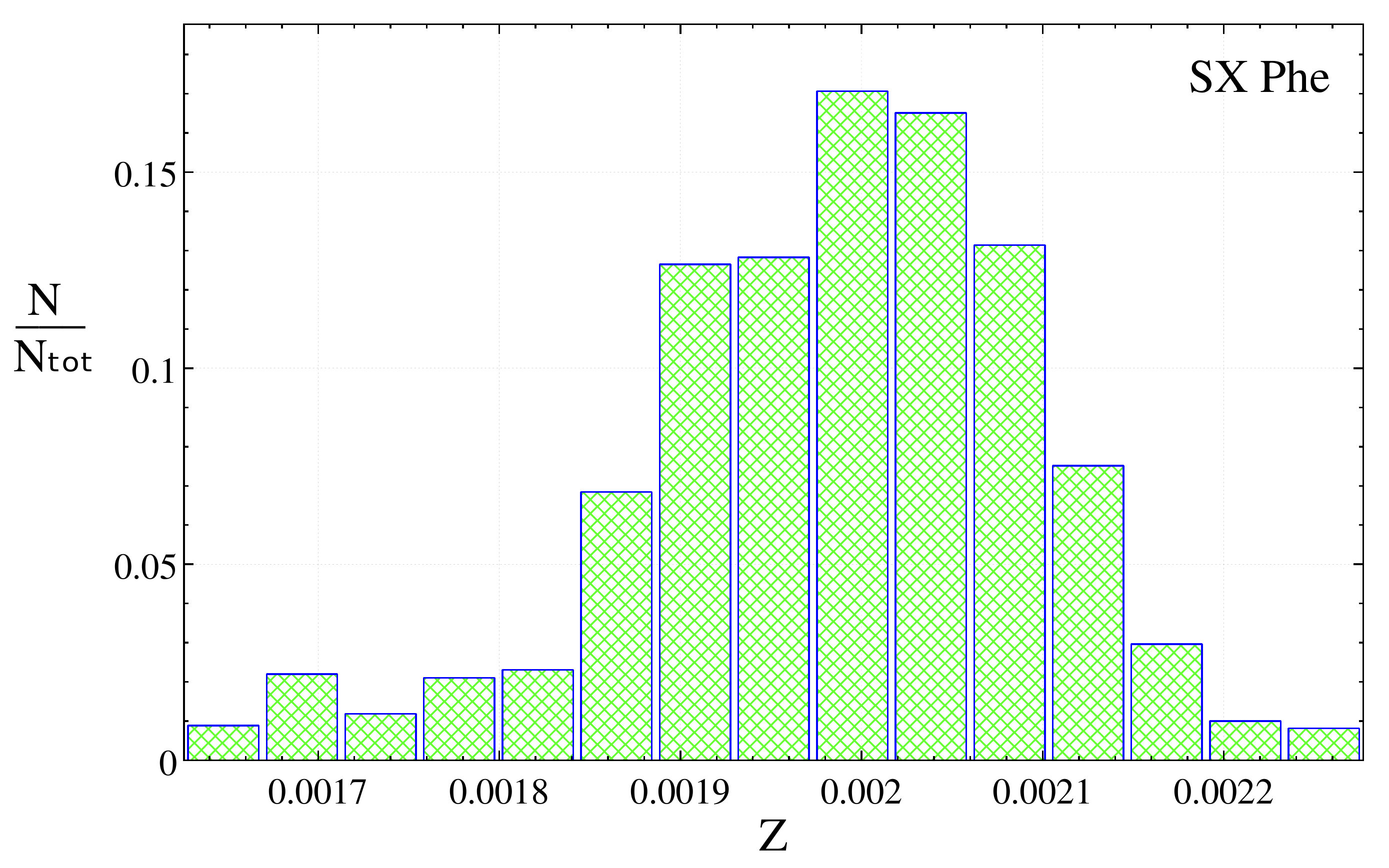}
	\caption{The normalized histograms for the metallicity $Z$ for  seismic models  computed with the OPAL opacities.}
	% of  BP Peg, AE UMa, RV Ari and SX Phe.}
\end{figure*}
\begin{figure*}
	\centering
	\includegraphics[clip,width=0.49\linewidth,height=63mm]{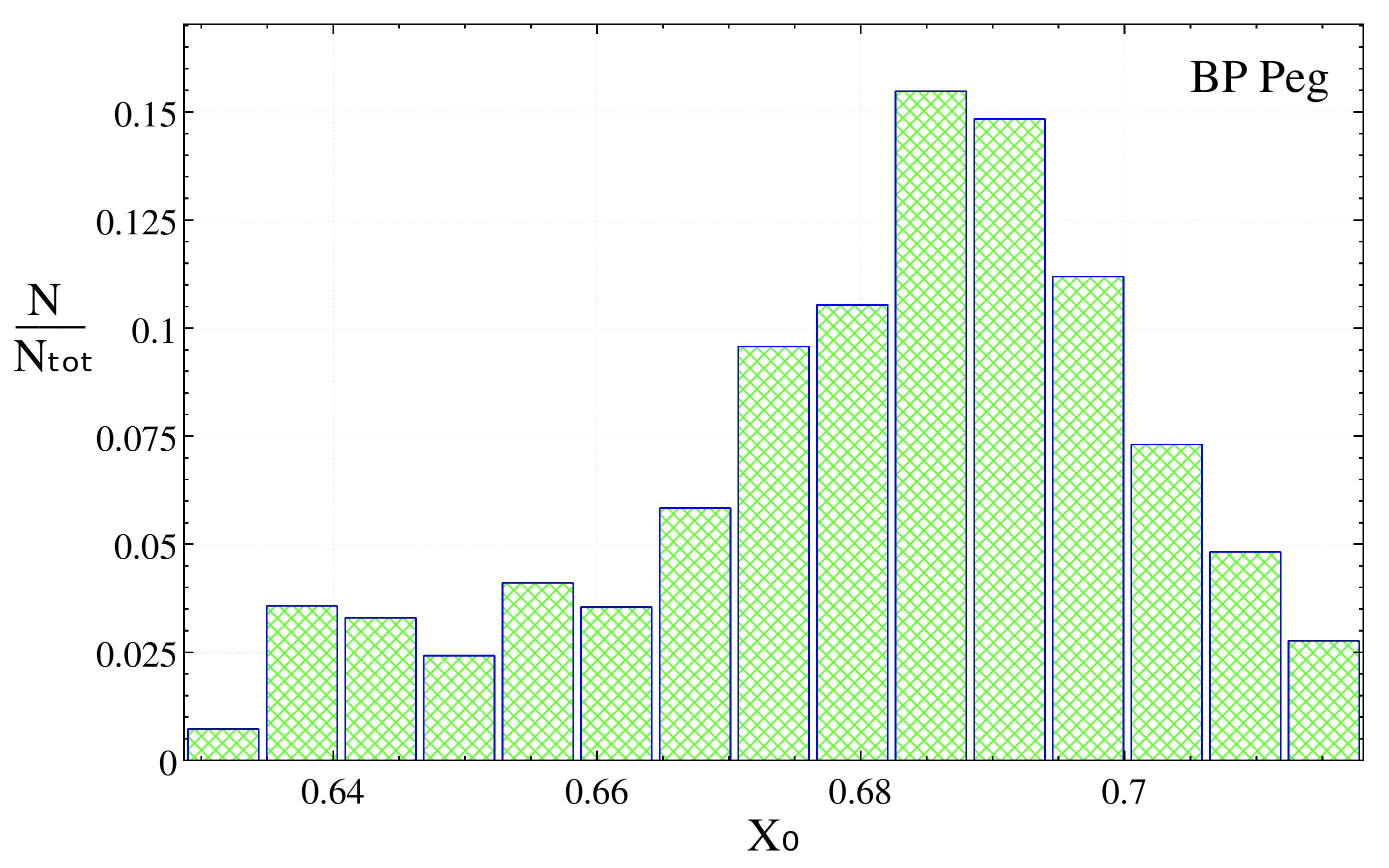}
	\includegraphics[clip,width=0.49\linewidth,height=63mm]{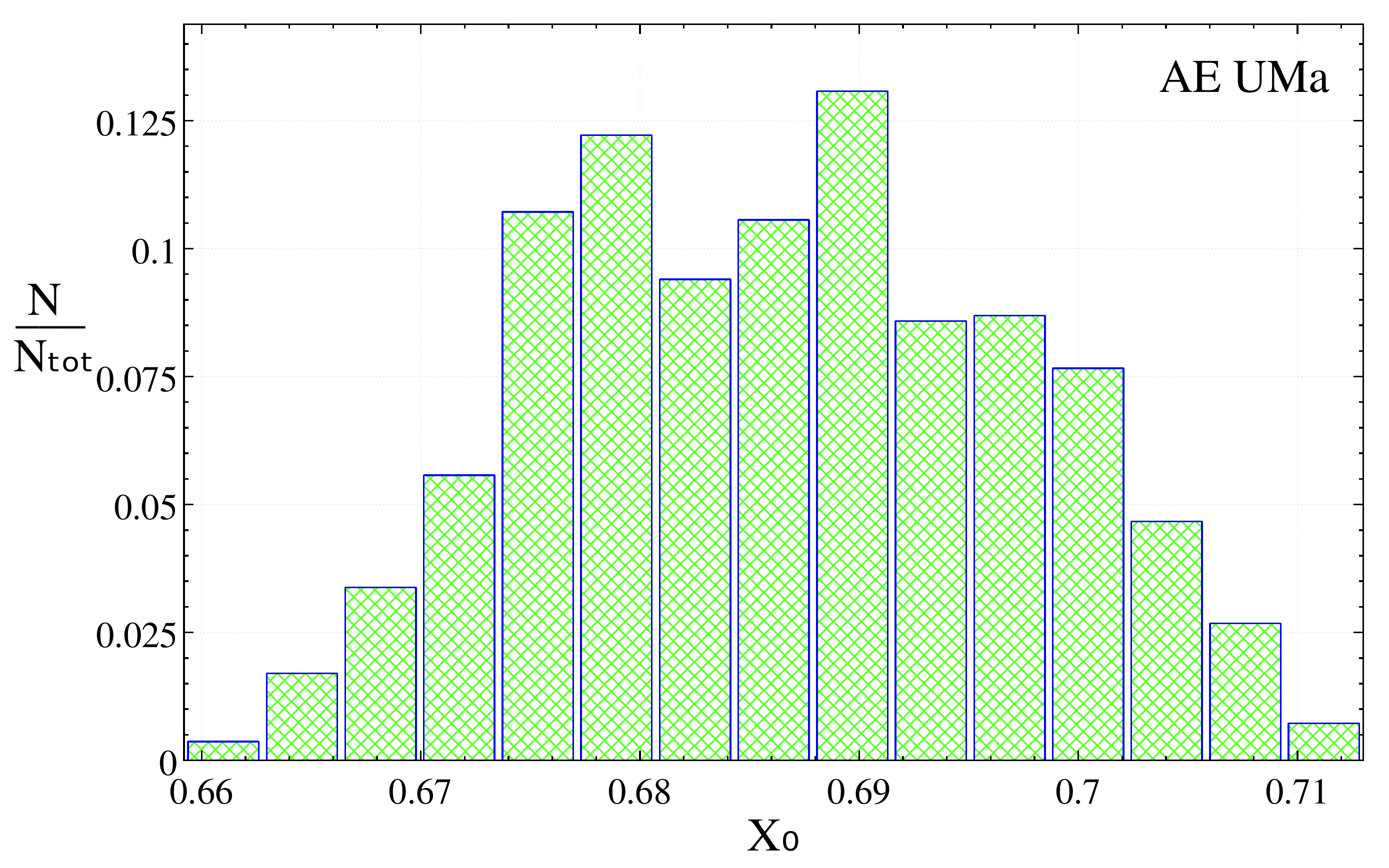}
	\includegraphics[clip,width=0.49\linewidth,height=63mm]{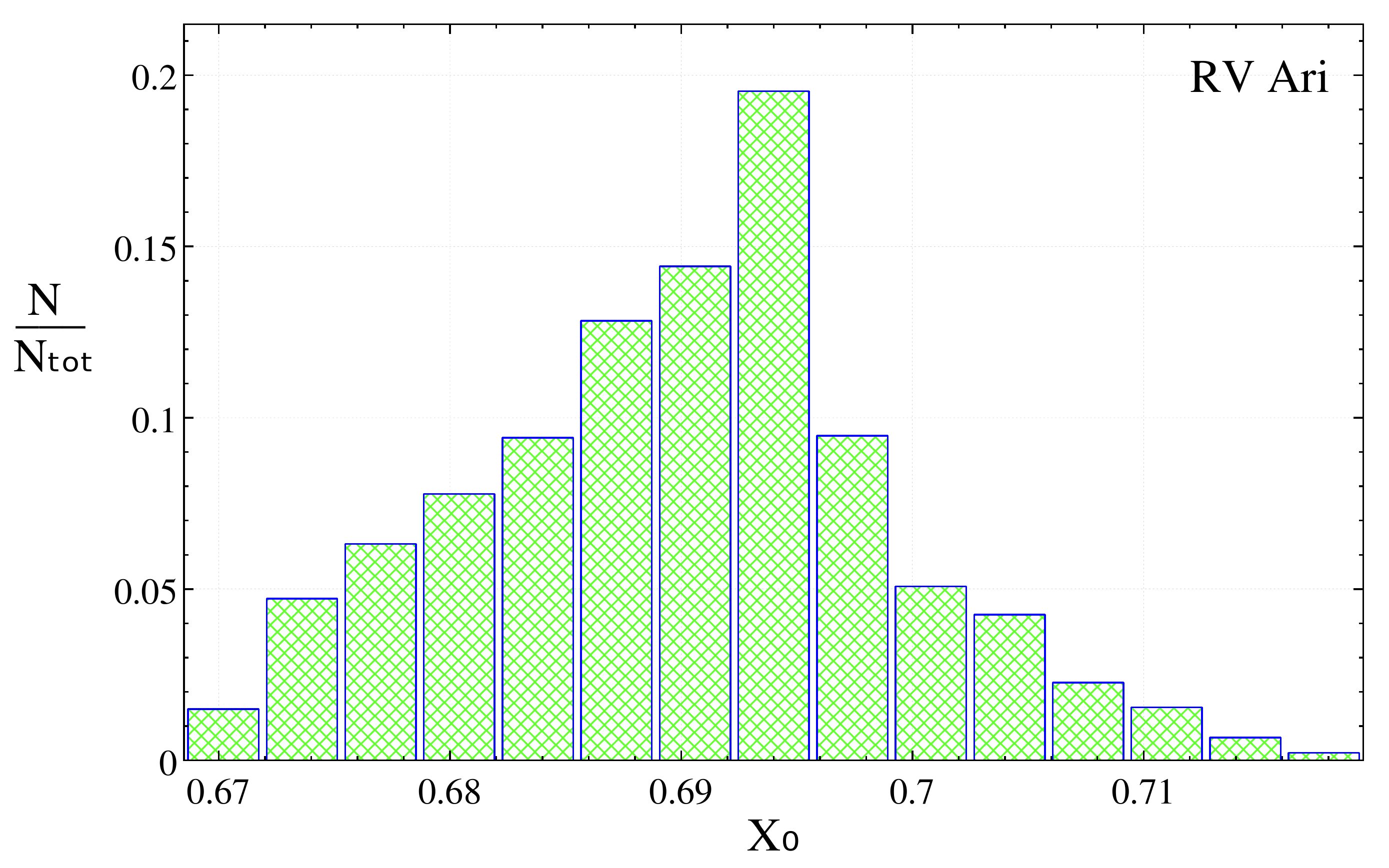}
	\includegraphics[clip,width=0.49\linewidth,height=63mm]{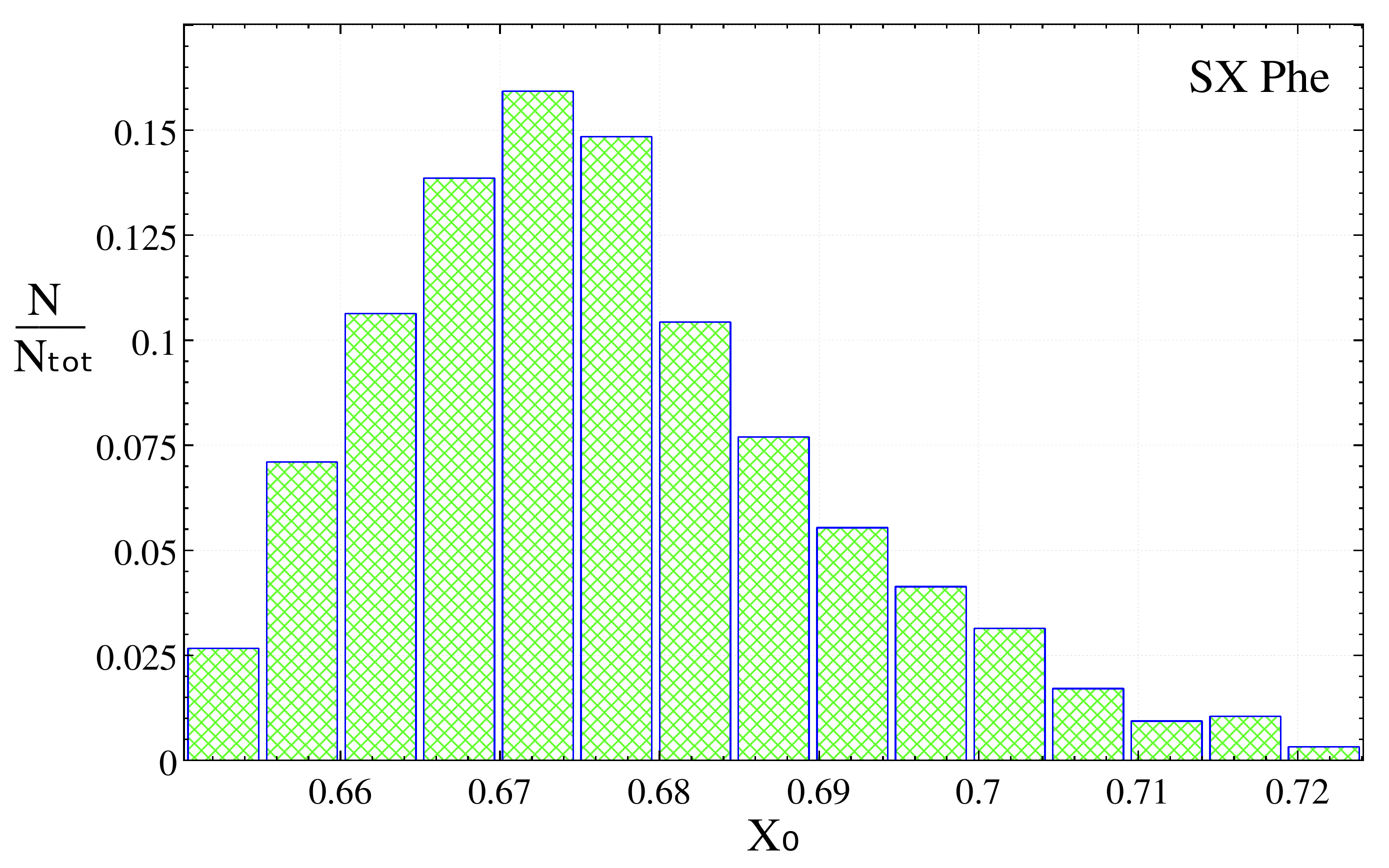}
	\caption{The normalized histograms for the initial hydrogen abundance $X_0$ for seismic models  computed with the OPAL opacities.}
	% of  BP Peg, AE UMa, RV Ari and SX Phe.}
\end{figure*}
\begin{figure*}
	\centering
	\includegraphics[clip,width=0.49\linewidth,height=63mm]{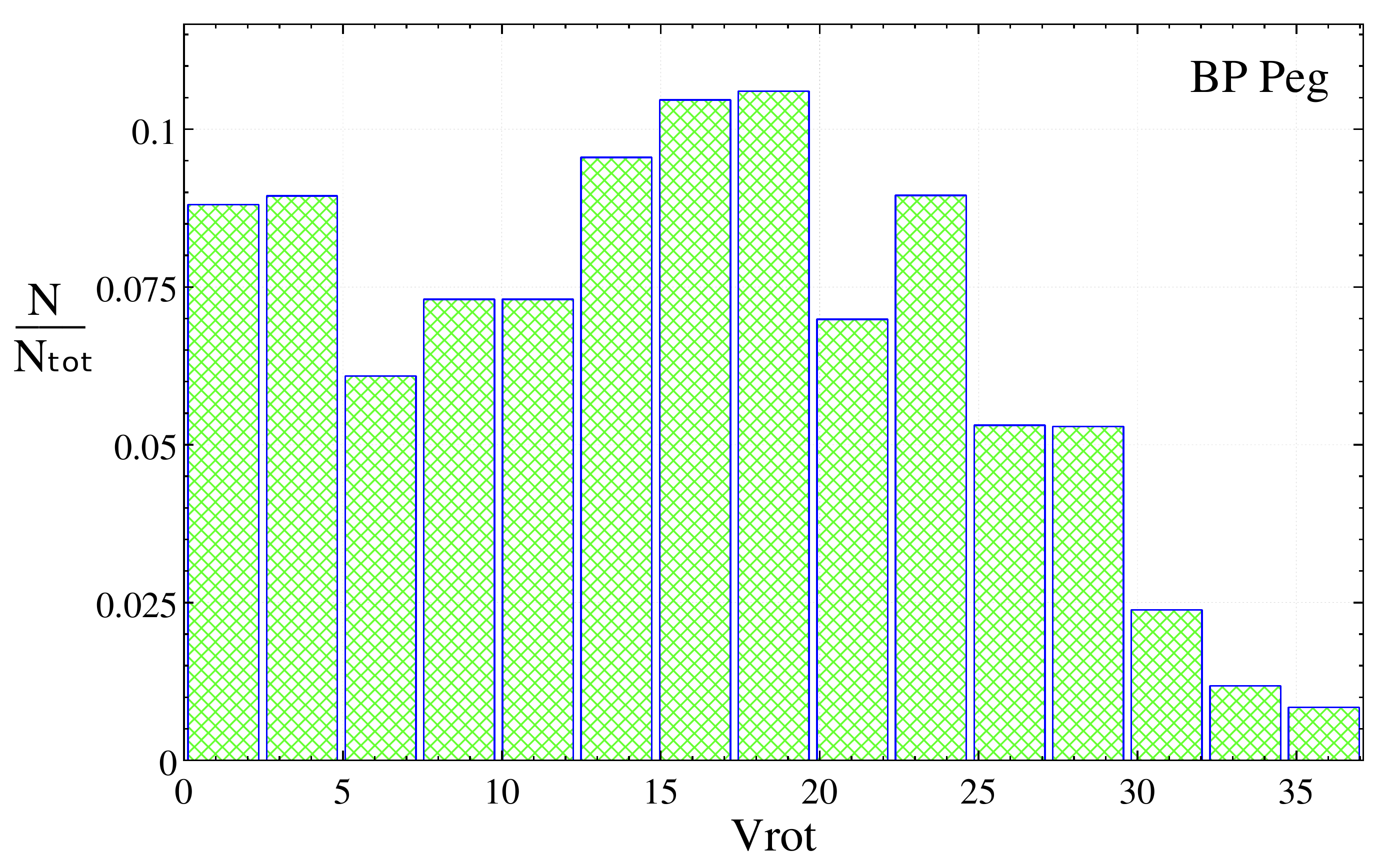}
	\includegraphics[clip,width=0.49\linewidth,height=63mm]{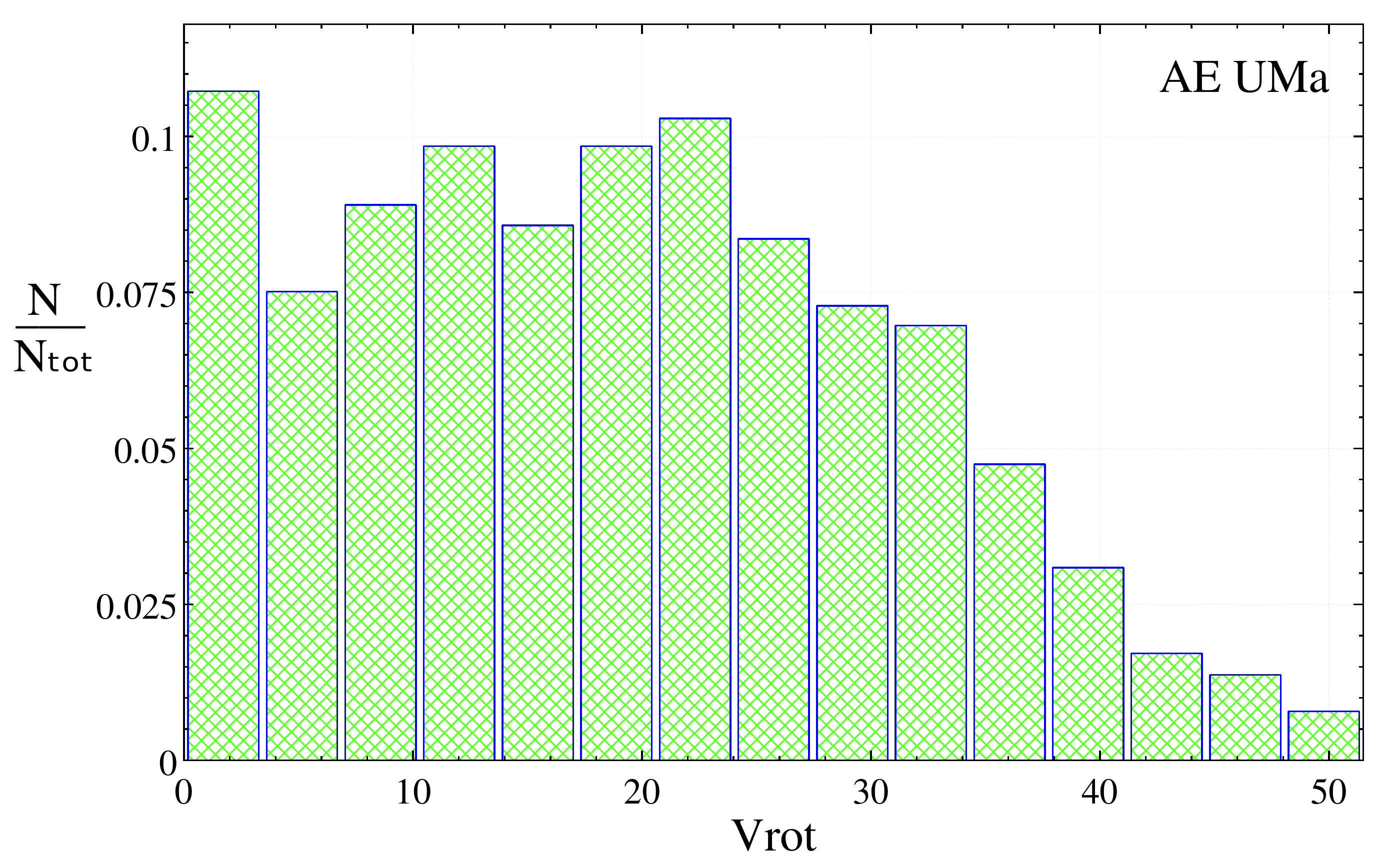}
	\includegraphics[clip,width=0.49\linewidth,height=63mm]{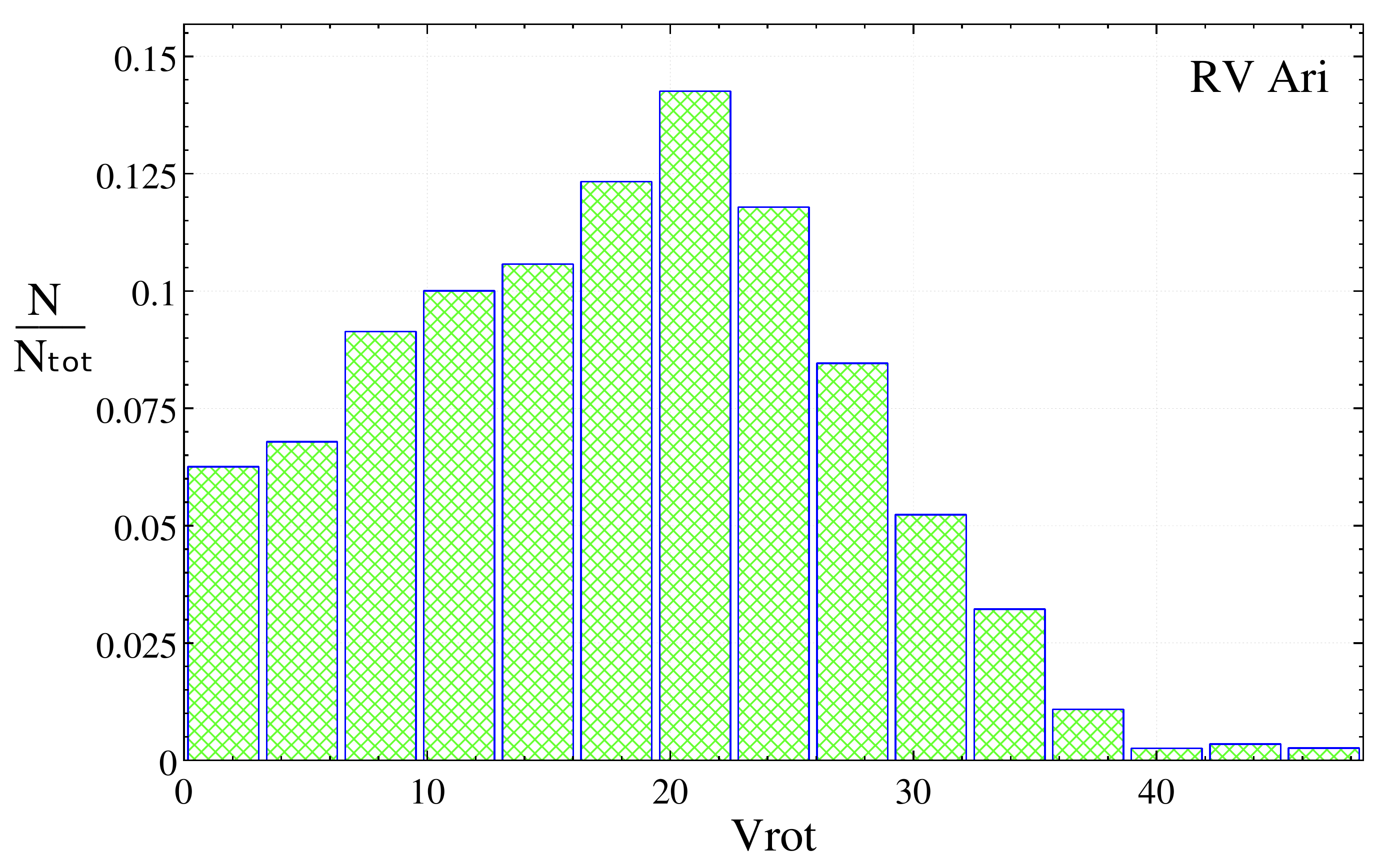}
	\includegraphics[clip,width=0.49\linewidth,height=63mm]{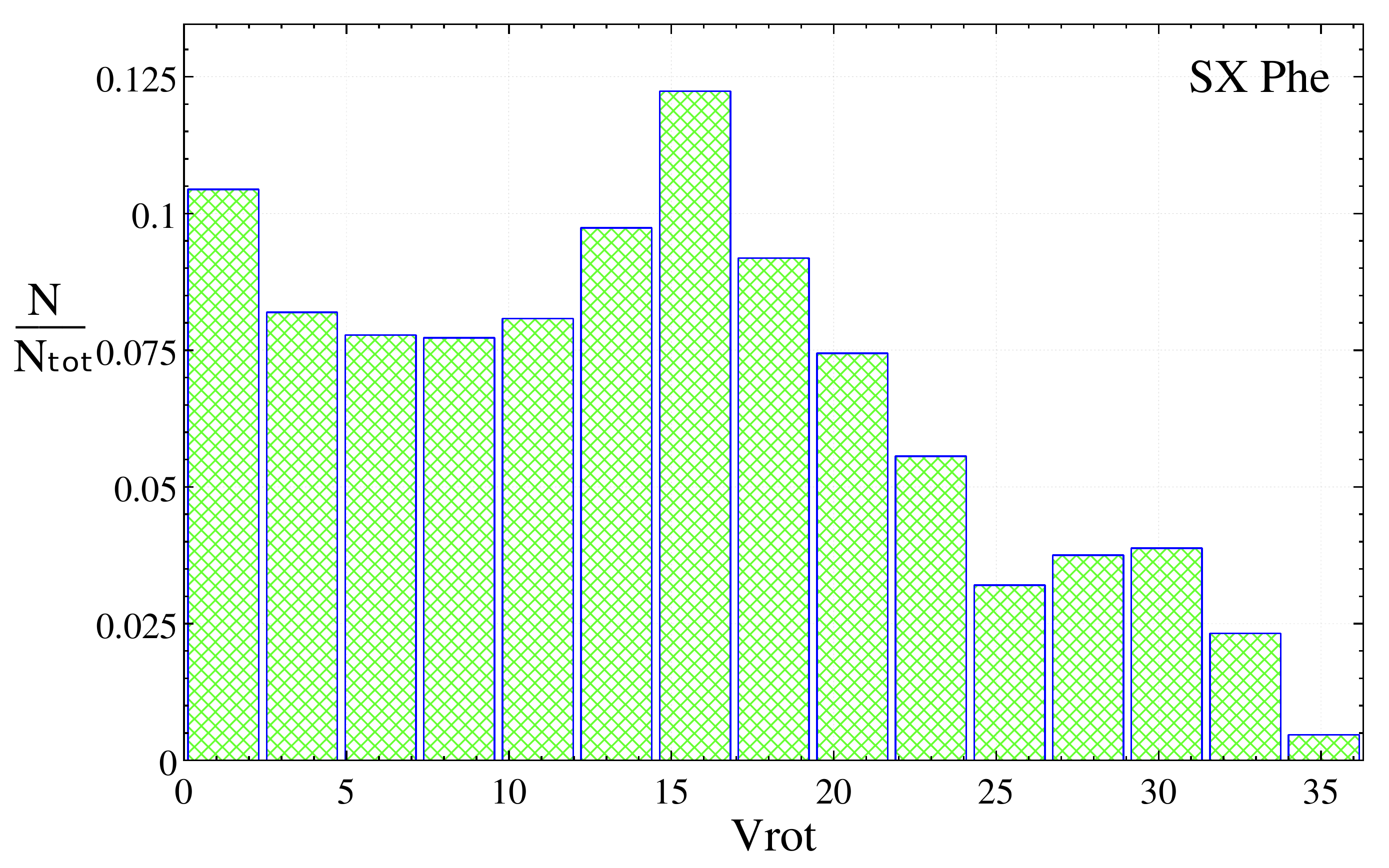}
	\caption{The normalized histograms for the current rotational velocity $V_{\rm rot}$ for seismic models  computed with the OPAL opacities.}
	% of  BP Peg, AE UMa, RV Ari and SX Phe.}
\end{figure*}
\begin{figure*}
	\centering
	\includegraphics[clip,width=0.49\linewidth,height=63mm]{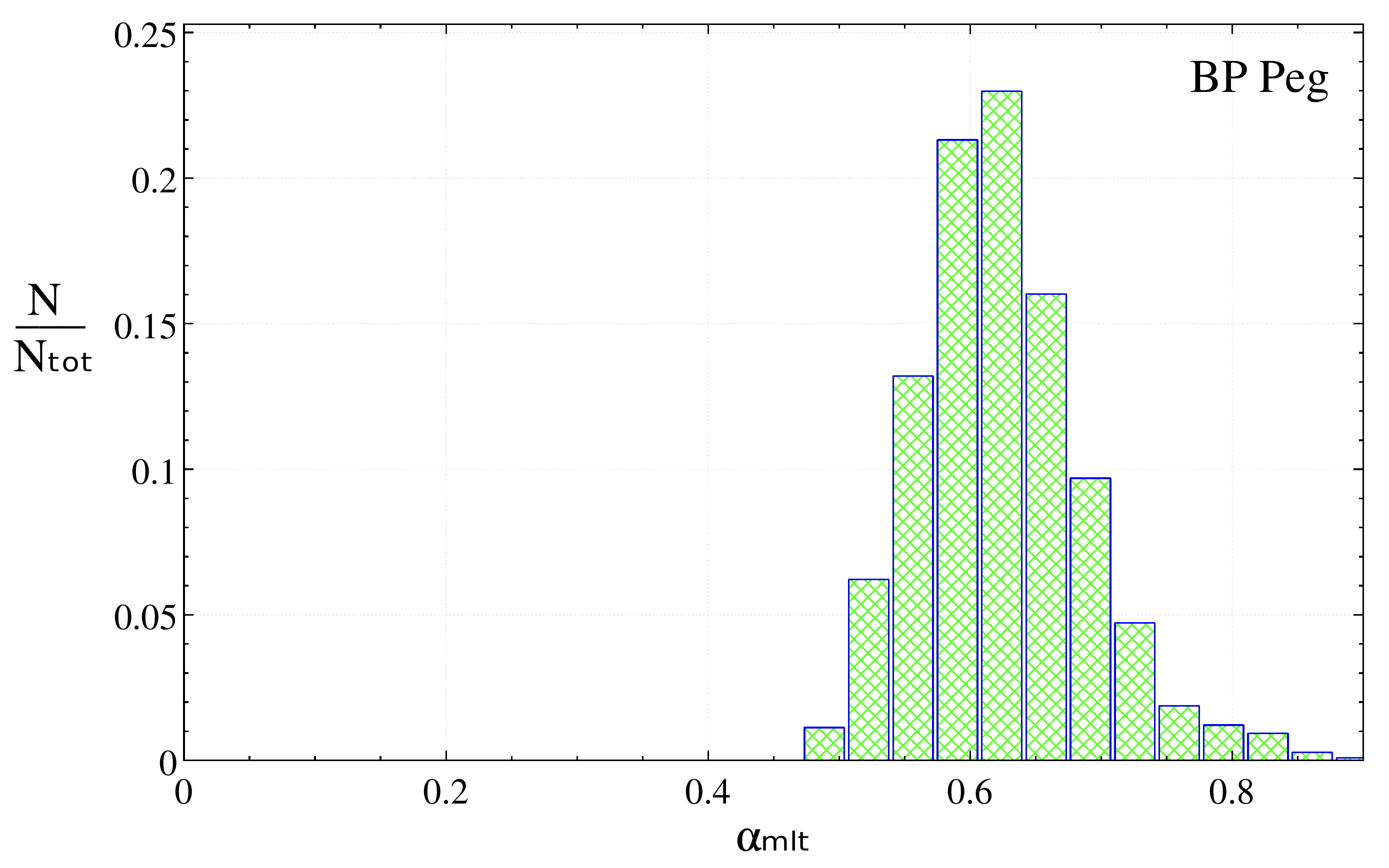}
	\includegraphics[clip,width=0.49\linewidth,height=63mm]{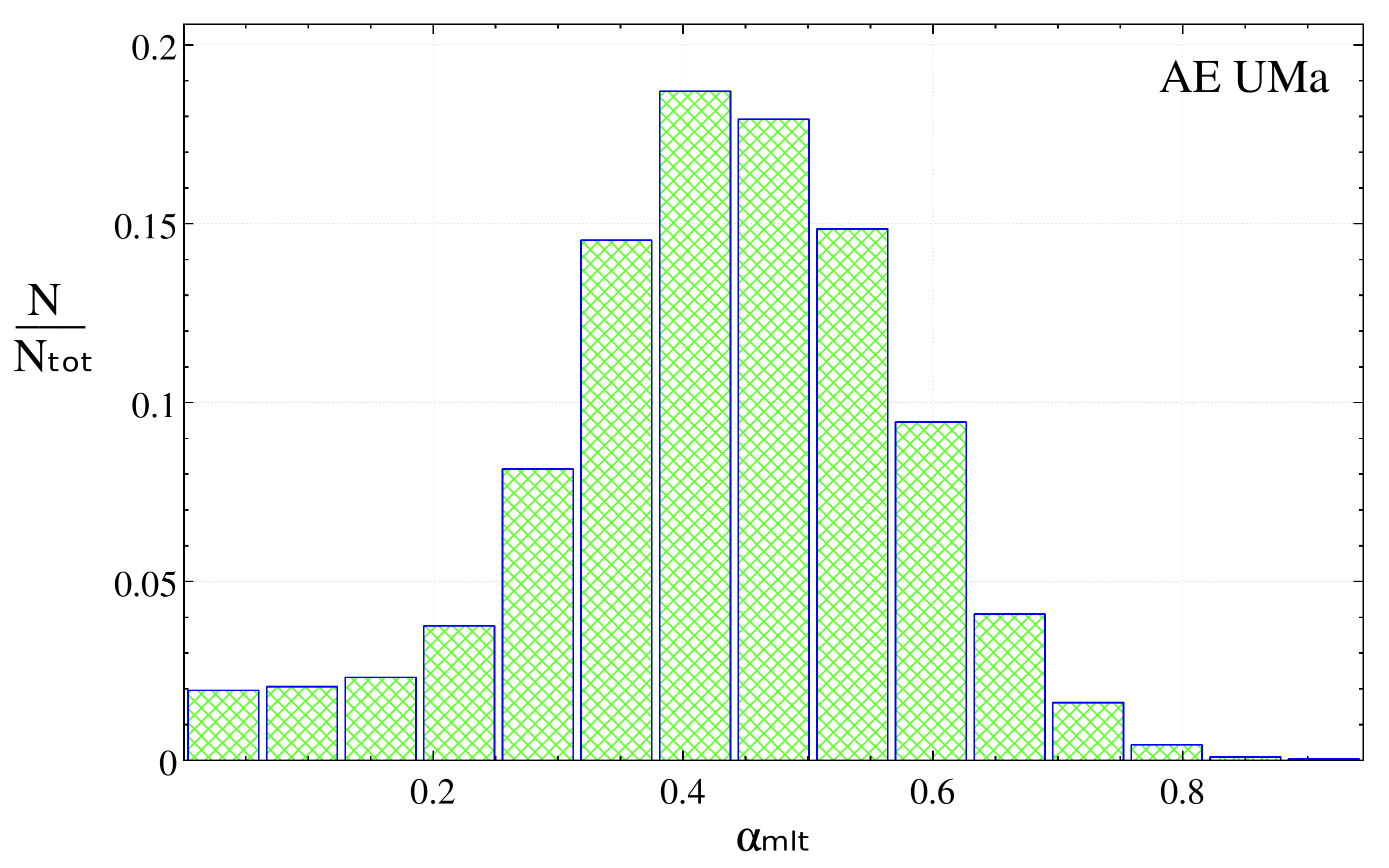}
	\includegraphics[clip,width=0.49\linewidth,height=63mm]{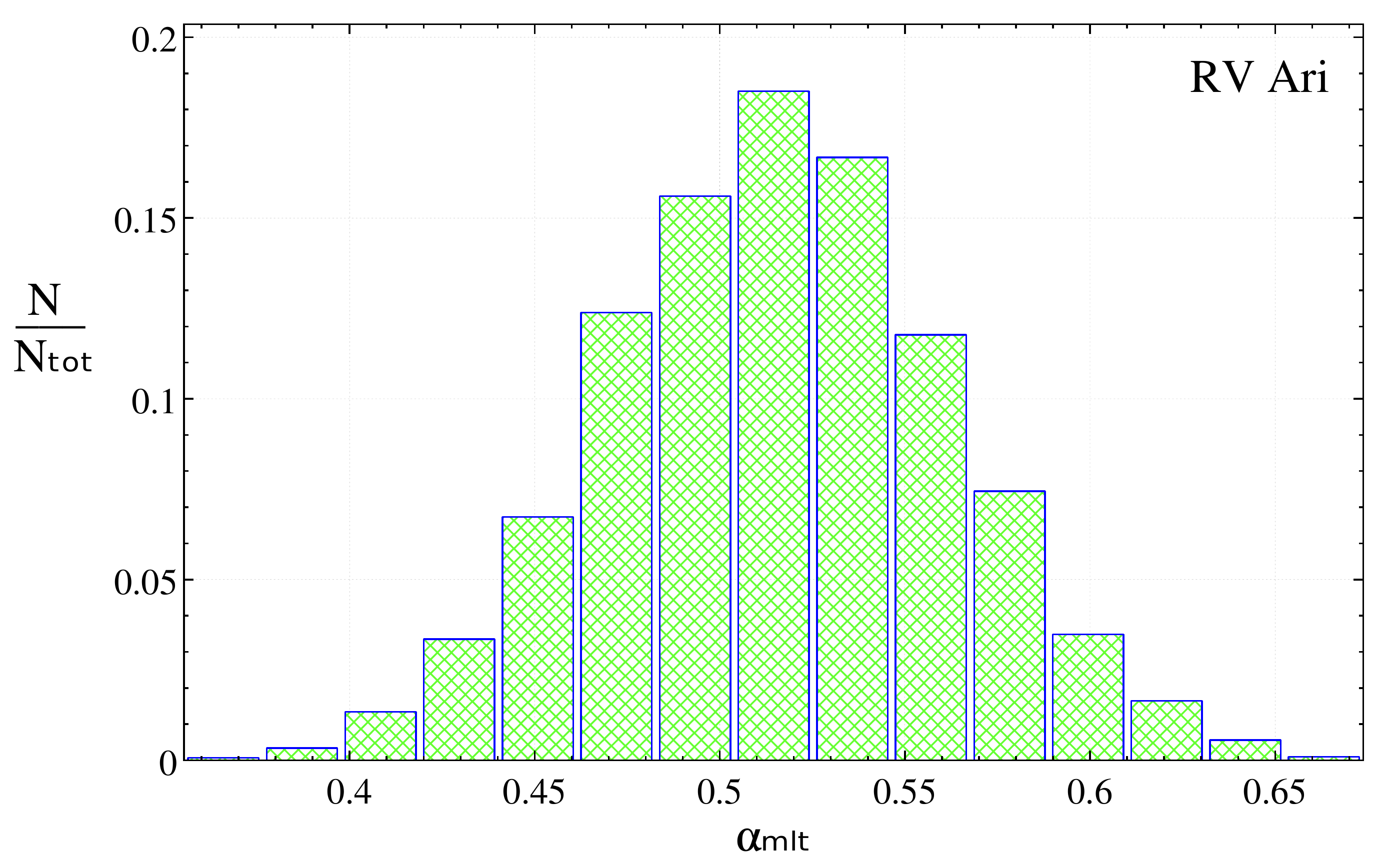}
	\includegraphics[clip,width=0.49\linewidth,height=63mm]{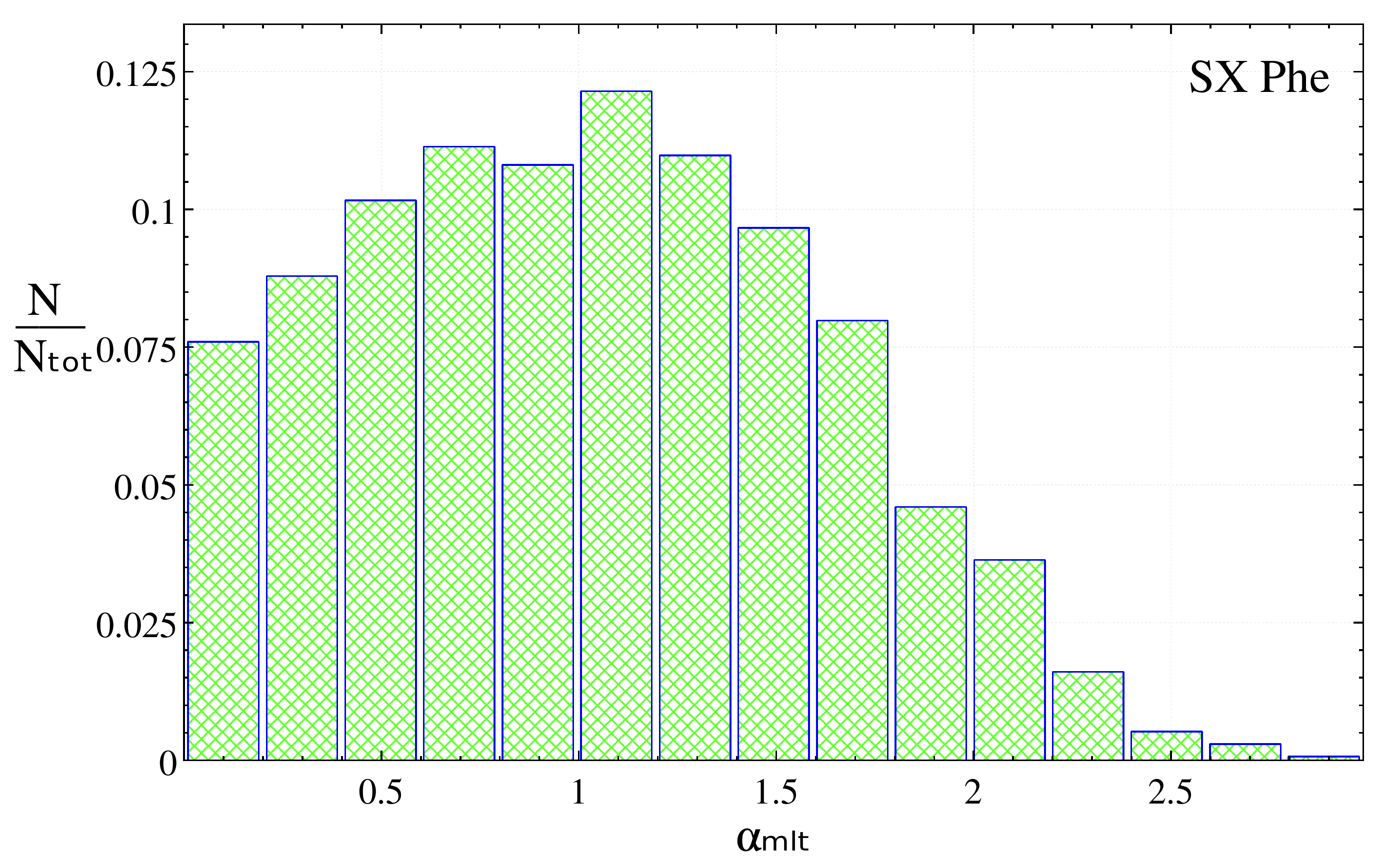}
	\caption{The normalized histograms for the mixing length parameter $\alpha_{\rm MLT}$ for  seismic models  computed with the OPAL opacities.}
	% of  BP Peg, AE UMa, RV Ari and SX Phe.}
\end{figure*}
 \begin{table*}
%\centering
\setlength{\tabcolsep}{3pt}
\caption{The expected values of the parameters of the four studied $\delta$ Sct stars from the Monte-Carlo simulations. The uncertainties were calculated as the square roots of the variance. Seismic models were computed with the OPAL opacities.}
	\begin{tabular}{ccccccccc }
		\hline
	star  & $M$  & $Z$    &  $X_0$  & $\alpha_{\rm MLT}$ &  $V_{\rm rot,0}$ & $V_{\rm rot}$ & $\log (T_{\rm eff}/{\rm K})$ & $\log L/$L$_\odot$ \\
	 &[M$_\odot$] &&&& [km$\cdot$s$^{-1}$] & [km$\cdot$s$^{-1}$]  & \\
\hline		
BP Peg&   1.81(4) & 0.0281(25) & 0.682(19) &     0.63(7)     &  16.0(9.3) & 15.1(8.8)      & 3.8353(16)     & 1.161(10) \\
 &&&&&&&& \\
AE UMa &  1.55(3)  & 0.0136(7) & 0.687(11)  & 0.43(14)  &  19.2(11.9) & 19.2(11.9)  & 3.8608(36)  & 1.085(18)  \\
 &&&&&&&& \\
RV Ari & 1.62(3)  & 0.0168(9) & 0.690(9)  & 0.52(5)  &  18.(9.3) & 17.6(9.2)  & 3.8494(22)  & 1.096(12) \\
 &&&&&&&& \\
SX Phe &  1.088(21) & 0.00199(5) & 0.677(14) & 1.04(58)  &  12.1(7.5) & 14.1(8.7) & 3.8986(37) & 0.889(17)\\
	\hline
	\end{tabular}
\end{table*}
\begin{table*}
\centering
\setlength{\tabcolsep}{3pt}
\caption{The median values of the parameters of the four studied $\delta$ Sct stars from the Monte-Carlo simulations. The uncertainties were calculated from quantiles  0.84 and 0.16. Seismic models were computed with the OPAL opacities.}
	\begin{tabular}{cccccccccc }
		\hline
	star  &   $M$    & $Z$    &  $X_0$  & $\alpha_{\rm MLT}$ &  $V_{\rm rot,0}$ & $V_{\rm rot}$ & $\log (T_{\rm eff}/{\rm K})$ & $\log L/$L$_\odot$ \\
	 &[M$_\odot$] &&&& [km$\cdot$s$^{-1}$] & [km$\cdot$s$^{-1}$] & \\
		\hline
		 &&&&&&&& \\
BP Peg&  $1.81^{+0.03}_{-0.04}$ &    $0.0271^{+0.0028}_{-0.0018}$ & $0.682^{+0.015}_{-0.023}$ &    $0.60^{+0.07}_{-0.06}$ &    $14.9^{+9.8}_{-11.5}$ & $14.1^{+9.1}_{-10.9}$ &  $3.8351^{+0.0013}_{-0.0018}$ & $1.158^{+0.010}_{-0.009}$ \\
 &&&&&&&& \\
AE UMa & $1.54^{+0.03}_{-0.02}$ &    $0.0135^{+0.0006}_{-0.0008}$ & $0.685^{+0.012}_{-0.011}$ &    $0.40^{+0.13}_{-0.14}$ &  $17.0^{+13.4}_{-12.9}$ & $17.0^{+13.5}_{-12.9}$ & $3.8606^{+0.0028}_{-0.0043}$ &    $1.082^{+0.017}_{-0.019}$ \\
 &&&&&&&& \\
RV Ari & $1.62^{+0.03}_{-0.03}$ &    $0.0164^{+0.0013}_{-0.0007}$ & $0.689^{+0.008}_{-0.010}$ &    $0.51^{+0.05}_{-0.05}$ &  $16.8^{+9.1}_{-10.9}$ & $16.4^{+9.0}_{-10.6}$ &  $3.8489^{+0.0020}_{-0.0023}$ & $1.092^{+0.013}_{-0.010}$\\
 &&&&&&&& \\
SX Phe & $1.083^{+0.024}_{-0.020}$ &   $0.00197^{+0.00010}_{-0.00011}$ & $0.672^{+0.016}_{-0.012}$ &  $0.92^{+0.64}_{-0.63}$ & $11.0^{+7.8}_{-8.5}$ & $12.8^{+9.1}_{-10.0}$ & $3.8973^{+0.0040}_{-0.0032}$ &    $0.884^{+0.019}_{-0.014}$ \\
 &&&&&&&& \\
		\hline
	\end{tabular}
\end{table*}

\end{document}